\documentclass[aps,prd,preprintnumbers,groupedaddress,nofootinbib,amssymb,notitlepage,eqsecnum]{revtex4-2}
\usepackage{here}
\usepackage[dvipdfmx]{graphicx}
\usepackage{amsmath,amsthm,amssymb}
\usepackage{bm}
\usepackage{color}

\usepackage{amsfonts}
\usepackage{dcolumn}
\usepackage{hyperref}
\allowdisplaybreaks[1]
\usepackage{stackengine}

\newcommand{\be}{\begin{equation}}  
\newcommand{\ee}{\end{equation}}
\newcommand{\ba}{\begin{eqnarray}}
\newcommand{\ea}{\end{eqnarray}}

\newcommand{\rd}{{\rm d}}

\newcommand{\bem}{\begin{bmatrix}}
\newcommand{\eem}{\end{bmatrix}}
\newcommand{\Mpl}{M_{\rm Pl}}

\allowdisplaybreaks

\begin{document}

\preprint{YITP-24-27, IPMU24-0006, WUCG-24-02}

\title{Angular and radial stabilities of 
spontaneously scalarized black holes \\
in the presence of scalar-Gauss-Bonnet couplings}

\author{Masato Minamitsuji$^{1}$\footnote{{\tt masato.minamitsuji@ist.utl.pt}}}
\author{Shinji Mukohyama$^{2,3}$\footnote{{\tt shinji.mukohyama@yukawa.kyoto-u.ac.jp}}}
\author{Shinji Tsujikawa$^{4}$\footnote{{\tt tsujikawa@waseda.jp}}}

\affiliation{
$^1$Centro de Astrof\'{\i}sica e Gravita\c c\~ao - CENTRA, Departamento de F\'{\i}sica, Instituto Superior T\'ecnico - IST, Universidade de Lisboa - UL, Av.~Rovisco Pais 1, 1049-001 Lisboa, Portugal}

\affiliation{
$^2$Center for Gravitational Physics and Quantum Information,
Yukawa Institute for Theoretical Physics, Kyoto University, 606-8502, Kyoto, Japan}

\affiliation{
$^3$Kavli Institute for the Physics and Mathematics of the Universe, the University of Tokyo Institutes for Advanced Study, the University of Tokyo, Kashiwa, Chiba, 277-8583, Japan}

\affiliation{
$^4$Department of Physics, Waseda University, 
3-4-1 Okubo, Shinjuku, Tokyo 169-8555, Japan}

\begin{abstract}

We study the linear stability of spontaneously scalarized black holes (BHs) induced by a scalar field $\phi$ coupled 
to a Gauss-Bonnet (GB) invariant $R_{\rm GB}^2$. For the scalar-GB coupling $\xi(\phi)=(\eta/8) (\phi^2+\alpha \phi^4)$, where $\eta$ and $\alpha$ are constants, we first show that 
there are no angular Laplacian instabilities of even-parity perturbations 
far away from the horizon for large multipoles $l \gg 1$.
The deviation of angular propagation speeds from the speed of light 
is largest on the horizon, whose property can be used to put constraints on the model parameters. 
For $\alpha \gtrsim -1$, the region in which the scalarized BH is subject to angular Laplacian instabilities 
can emerge. Provided that $\alpha \lesssim -1$ and $-1/2<\alpha \phi_0^2<-0.1155$, where $\phi_0$ is the field value on the horizon with a unit of the reduced Planck mass $\Mpl=1$, there are scalarized BH solutions satisfying  all the linear stability conditions throughout the horizon exterior.
We also study the stability of spontaneously scalarized BHs in scalar-GB theories with a nonminimal coupling $-\beta \phi^2 R/16$, where $\beta$ is a positive constant and $R$ is a Ricci scalar.
As the amplitude of the field on the horizon approaches an upper limit $|\phi_0|=4/\sqrt{\beta}$, one of the squared angular propagation speeds $c_{\Omega-}^2$ enters the instability region $c_{\Omega-}^2<0$. So long as $|\phi_0|$ is smaller than a maximum value determined for each $\beta$ in the range $\beta>5$, however, the scalarized BHs are linearly stable in both angular and radial directions.

\end{abstract}

\date{\today}


\maketitle

\section{Introduction}
\label{introsec}

In General Relativity (GR), the Schwarzschild black hole (BH) arises as a unique vacuum solution of the Einstein equation on a static and spherically symmetric background. 
The geometry of the Schwarzschild BH is solely characterized by a BH mass. In order to generate an additional ``hair'' to the BH mass on the same background, we need to take into account 
new degrees of freedom such as scalar and vector fields. 
In the presence of a $U(1)$ gauge-invariant vector field, 
the resulting solution is the Reissner-Nordstr\"om (RN) BH with an electric and/or magnetic charge besides the mass. 

If we consider a canonical scalar field $\phi$ with the potential $V(\phi)$ satisfying certain conditions, it is known that static, spherically symmetric, and asymptotically flat
BHs do not acquire new scalar hairs \cite{Hawking:1971vc,Bekenstein:1972ny,Herdeiro:2015waa}. 
This no-hair property also persists for k-essence \cite{Graham:2014mda} and for a nonminimally coupled 
scalar field with the Ricci scalar $R$ of the form $\kappa (\phi)R$ \cite{Hawking:1972qk,Bekenstein:1995un,Sotiriou:2011dz,Faraoni:2017ock,Hui:2012qt}.
A simple way to evade the no-hair theorems is to introduce 
time-dependence to the scalar profile while keeping the time-independent metric\footnote{\label{stealthsol}This allows for, 
for example, stealth BH solutions, in which the metric is the same as in GR but the scalar profile is non-trivial 
(see Ref.~\cite{Mukohyama:2005rw} for the first stealth Schwarzschild solution in k-essence). 
For each stealth BH, if and only if the scalar profile is timelike, one can easily implement the so-called scordatura mechanism~\cite{Arkani-Hamed:2003pdi,Motohashi:2019ymr,DeFelice:2022xvq} 
to find the corresponding approximately stealth solution (see 
Ref.~\cite{Mukohyama:2005rw} for the first example), which is stealth for any practical purposes at astrophysical scales~\cite{Mukohyama:2005rw,Cheng:2006us,DeFelice:2022qaz} and around which perturbations are weakly coupled all the way up to the scale associated with the background value of the derivative of the scalar field~\cite{Arkani-Hamed:2003pdi}.}. In this paper, however, we shall not pursue this possibility and we assume that 
scalar-field profiles for the background solutions are 
time-independent. 

In the presence of a coupling between $\phi$ and the Gauss-Bonnet (GB) invariant $R_{\rm GB}^2$ of the form $\xi(\phi)R_{\rm GB}^2$, there are hairy BH solutions endowed with scalar hairs \cite{Kanti:1995vq,Torii:1996yi,Kanti:1997br,Chen:2006ge,Guo:2008hf,Guo:2008eq,Pani:2009wy,Ayzenberg:2014aka,Maselli:2015tta,Kleihaus:2011tg,Kleihaus:2015aje,Sotiriou:2013qea,Sotiriou:2014pfa,Tsujikawa:2023egy}. 
Such Einstein-scalar-Gauss-Bonnet (EsGB) theories belong to a subclass of Horndeski theories with second-order field equations of motion \cite{Horndeski:1974wa,Deffayet:2011gz,Kobayashi:2011nu,Charmousis:2011bf}. 
Indeed, even in the framework of full Horndeski and DHOST theories, it was shown that the existence of the scalar-GB coupling is necessary for realizing non-pathological (e.g., linearly stable) and asymptotically flat BH solutions with a regular scalar field on the horizon \cite{Creminelli:2020lxn,Minamitsuji:2022vbi} (see also Ref.~\cite{Minamitsuji:2022mlv}).

In EsGB theories given by the coupling $\xi(\phi)R_{\rm GB}^2$, it is known that a phenomenon called spontaneous scalarization of BHs and 
neutron stars (NSs) can occur for $\xi(\phi)$ containing even power-law functions of $\phi$ \cite{Doneva:2017bvd,Silva:2017uqg,Doneva:2017duq,Silva:2018qhn,Antoniou:2017acq,Antoniou:2017hxj,Minamitsuji:2018xde,Doneva:2018rou,Cunha:2019dwb,Brihaye:2019kvj,Dima:2020yac,Herdeiro:2020wei,Berti:2020kgk,Doneva:2021dqn}. 
This is analogous to spontaneous scalarization of NSs induced 
by the nonminimal coupling $\kappa (\phi)R$ \cite{Damour:1993hw,Damour:1996ke}.
In the latter case, the trace $T$ of matter inside the star is coupled to the scalar field $\phi$ through a relation $\Mpl^2 R=-T$. 
For BHs, the Ricci scalar is vanishing due to the absence of matter ($T=0$).
However, the GB invariant does not vanish for a vacuum Schwarzschild or 
a Kerr background. Hence a BH can acquire a nontrivial scalar hair through 
tachyonic instability caused by the coupling between 
$\phi$ and $R_{\rm GB}^2$. 

For the sGB coupling given by even-power law functions like $\xi(\phi)=\eta \phi^2/8$ with $\eta$ being constant, there is a nonvanishing scalar field branch $\phi \neq 0$ besides the GR branch $\phi=0$. 
On weak gravitational backgrounds like the Solar System, the scalar field stays in the latter trivial branch, whose property does not conflict 
with the experimental tests on gravity \cite{Will:2014kxa}. 
In the vicinity of BHs, the effective mass squared of $\phi$, which is given by $m_{\rm eff}^2=-\xi_{,\phi \phi} R_{\rm GB}^2$, 
can be negative at $\phi=0$ under the condition 
$\xi_{,\phi \phi} \equiv {\rm d}^2 \xi/{\rm d}\phi^2>0$. 
For the coupling $\xi(\phi)=\eta \phi^2/8$ this condition translates to $\eta>0$, under which the GR branch can 
trigger tachyonic instability toward the other nontrivial branch to realize scalarized BHs (see also 
Refs.~\cite{Myung:2020etf,Doneva:2021dcc,Herdeiro:2018wub,Myung:2018vug,Fernandes:2019rez,Fernandes:2019kmh,Ikeda:2019okp,Hod:2019ulh,Minamitsuji:2021vdb,Minamitsuji:2022qku} 
for spontaneous scalarization caused by other couplings).

In EsGB theories given by the coupling $\xi(\phi)=\eta \phi^2/8$, however, 
the scalarized branch ($\phi \neq 0$) is unstable against radial perturbations everywhere \cite{Blazquez-Salcedo:2018jnn}.
Taking into account higher-order terms to the sGB coupling, e.g., $\xi(\phi)=(\eta/8) ( \phi^2+\alpha \phi^4 )$ with $\alpha$ being  dimensionless coupling constant, there are parameter spaces in which the scalarized BH solutions are not subject to radial tachyonic instabilities \cite{Minamitsuji:2018xde,Silva:2018qhn,Macedo:2019sem}. 
In this paper we will study all the linear stability conditions for scalarized BHs with the sGB coupling $\xi(\phi)=(\eta/8) ( \phi^2+\alpha \phi^4)$ in both angular and radial directions by exploiting linear stability conditions derived for 
Horndeski theories \cite{Kase:2021mix} (see also 
Refs.~\cite{Kobayashi:2012kh,Kobayashi:2014wsa,Kase:2023kvq}). 
Indeed, there is a wide range of parameter spaces in which 
the scalarized BHs are subject to neither ghost nor Laplacian 
instabilities for large radial and angular momentum modes.

The BH scalarization model in EsGB theories was extended to 
include the nonminimal coupling $\kappa (\phi)R$, 
where $\kappa (\phi)$ is an even power-law function of $\phi$ \cite{Antoniou:2020nax,Antoniou:2021zoy,Antoniou:2022agj}. 
Such Einstein-scalar-Gauss-Bonnet-Ricci (EsGBR) theories were 
motivated for realizing the asymptotic GR solution 
$\phi=0$ as a cosmological attractor \cite{Antoniou:2020nax}. 
For the couplings $\kappa (\phi)=-\beta \phi^2/16$ {with $\beta$ being the dimensionless constant and $\xi(\phi)=\eta \phi^2/8$, the analysis of Ref.~\cite{Antoniou:2022agj} showed that there are radially stable scalarized BH solutions if $\beta>5$. 

Recently, it was found that scalarized BHs in EsGBR theories can be subject to angular instabilities for the quadrupole 
perturbation ($l=2$) \cite{Kleihaus:2023zzs}. 
This stability analysis is based on the computation of the BH entropy \cite{Wald:1993nt} integrated over the spatial 
cross section of the horizon. 
However, it is not yet clear what happens for the angular stability 
of BHs in the large multipole limit ($l \gg 1$). 
In particular, since $l$ appears in the quadratic action for perturbations only through the expression $l(l+1)$ (which is an increasing function of $l$), if increasing $l$ from $0$ to $2$ changes stable perturbations to unstable ones as claimed 
in Ref.~\cite{Kleihaus:2023zzs}, then the instability is naturally expected to be more prominent for larger $l$. 
Since EsGBR theories belong to a subclass of Horndeski theories, one can exploit the angular stability conditions derived 
in Ref.~\cite{Kase:2021mix} for $l \gg 1$.
In this paper, we will address both the angular and radial stabilities of scalarized BHs in EsGBR theories and find that the angular Laplacian instabilities can arise for large field values $\phi_0$ on the horizon. 
However, we will show the existence of parameter spaces in which neither ghosts, angular/radial Laplacian instabilities, nor 
radial tachyonic instabilities are present.

This paper is organized as follows. 
In Sec.~\ref{classsec}, we briefly review conditions for tachyonic instabilities of the GR branch in both EsGB and EsGBR theories and also present the background equations of motion on the static and spherically symmetric background.
In Sec.~\ref{stabilitysec}, we revisit the linear stability conditions derived in \cite{Kase:2021mix} and apply them 
to EsGB and EsGBR theories in the small field regime ($|\phi| \ll 1$). 
In Sec.~\ref{radisec}, we study the radial stability of the dynamical 
scalar-field perturbation $\delta \phi$ by considering a monopole mode ($l=0$). 
After deriving a closed-form second-order differential equation for $\delta \phi$ with the backreaction of metric perturbations taken into account, we obtain the regime of model parameters consistent with the radial stability analysis performed in the literature \cite{Minamitsuji:2018xde,Antoniou:2020nax}.
In Sec.~\ref{angsec1} and \ref{angsec2}, we elucidate the parameter spaces of EsGB and EsGBR theories in which no linear instabilities of scalarized BHs are present. 
Sec.~\ref{consec} is devoted to conclusions.

\section{EsGB and EsGBR theories}
\label{classsec}

We begin with a general action of EsGBR theories given by 
\be
{\cal S}=\frac{1}{2} \int \rd^4 x \sqrt{-g} 
\left[ R+X+\xi(\phi) R_{\rm GB}^2+\kappa(\phi)R \right]\,,
\label{action}
\ee
where $g$ is the determinant of metric tensor $g_{\mu \nu}$, $R$ is the Ricci scalar, $X=-(1/2)g^{\mu \nu} \partial_{\mu}\phi \partial_{\nu}\phi$ is the kinetic term of a scalar field $\phi$, $R_{\rm GB}^2$ is the GB invariant, and $\xi(\phi)$ and $\kappa(\phi)$ are functions of $\phi$.
We will choose the unit of $\Mpl=1$, where $\Mpl$ represents 
the reduced Planck mass.
Note that EsGB theories correspond to the special case where $\kappa(\phi)=0$.

On the static and spherically symmetric background, spontaneous scalarization of BHs can be realized for the sGB coupling $\xi(\phi)$ with a $Z_2$ symmetry $\xi(-\phi)=\xi(\phi)$ \cite{Doneva:2017bvd,Silva:2017uqg,Silva:2018qhn,Antoniou:2017acq,Antoniou:2017hxj,Minamitsuji:2018xde}. 
If we apply these EsGB theories to cosmology, the asymptotic GR solution $\phi=0$ is not generally realized without tuning the initial conditions \cite{Anson:2019uto,Franchini:2019npi,Antoniou:2020nax}. 
If we allow the existence of a nonminimal coupling of the form $-\beta \phi^2 R/16$, the scalar field can approach a cosmological attractor being compatible with solar-system constraints on today's field value of $\phi$ \cite{Antoniou:2020nax,Antoniou:2021zoy,Antoniou:2022agj}.
In such EsGBR theories, it was recently recognized that scalarized BHs can be subject to a quadrupole angular instability below a critical value of the BH mass \cite{Kleihaus:2023zzs}. 
If this is the case, then the similar angular instability may also manifest itself for larger multipoles $l \gg 1$. 
In this paper, we will address this latter issue along with the radial stability by using general conditions derived in Ref.~\cite{Kase:2021mix}.
Moreover, we will investigate the angular and radial stabilities of BHs in EsGB theories. 

The action (\ref{action}) in EsGBR theories belongs to a subclass of 
Horndeski theories given by the action 
\ba
{\cal S} &=&
\int {\rm d}^4 x \sqrt{-g}\,
\bigg\{ G_2-G_{3}\square\phi 
+G_{4} R +G_{4,X} \left[ (\square \phi)^{2}
-(\nabla_{\mu}\nabla_{\nu} \phi)
(\nabla^{\mu}\nabla^{\nu} \phi) \right] \nonumber \\
& &+G_{5} G_{\mu \nu} \nabla^{\mu}\nabla^{\nu} \phi
-\frac{1}{6} G_{5,X}
\left[ (\square \phi )^{3}-3(\square \phi)\,
(\nabla_{\mu}\nabla_{\nu} \phi)
(\nabla^{\mu}\nabla^{\nu} \phi)
+2(\nabla^{\mu}\nabla_{\alpha} \phi)
(\nabla^{\alpha}\nabla_{\beta} \phi)
(\nabla^{\beta}\nabla_{\mu} \phi) \right]\bigg\},
\label{action2}
\ea
where $\nabla_{\mu}$ is a covariant derivative operator, $\square=\nabla^{\mu}\nabla_{\mu}$, $G_{\mu \nu}$ is the Einstein tensor, and 
\ba
G_2 &=& \frac{1}{2} \left[ 
X+8 \xi_{,\phi \phi \phi \phi}(\phi) X^2 
\left( 3-\ln |X| \right) \right]\,,\\
G_3 &=&  2 \xi_{,\phi \phi \phi}(\phi) X 
\left[ 7-3\ln |X| \right]\,,\\
G_4 &=& \frac{1}{2} \left[ 1+\kappa(\phi) \right]
+2 \xi_{,\phi \phi}(\phi) X \left( 2-\ln |X| \right)\,,\\
G_5 &=& -2  \xi_{,\phi}(\phi) \ln |X|\,,
\ea
with the notations 
$\xi_{,\phi}(\phi)=\rd \xi(\phi)/\rd \phi$, 
$\xi_{,\phi \phi}(\phi)=\rd^2 \xi(\phi)/\rd \phi^2$, etc.

Varying the action (\ref{action}) with respect to $\phi$, we obtain 
the scalar-field equation of motion 
\be
\square \phi+R_{\rm GB}^2\,\xi_{,\phi} 
+R\,\kappa_{,\phi}=0\,.
\label{phieq0}
\ee
For the functions $\xi(\phi)$ and $\kappa(\phi)$ respecting the 
$Z_2$ symmetry, there are in general the GR branch $\phi=0$ 
and the nonvanishing scalar-field branch $\phi \neq 0$. 

We consider the static and spherically symmetric background 
given by the line element 
\be
{\rm d}s^{2} =-f(r) {\rm d}t^{2} +h^{-1}(r) {\rm d}r^{2} + 
r^{2} \left( {\rm d}\theta^{2}
+\sin^{2}\theta\, {\rm d} \varphi^{2} \right)\,,
\label{spmetric}
\ee
where $t$, $r$ and $(\theta,\varphi)$ represent the time, 
radial, and angular coordinates 
(in the ranges $0 \le \theta \le \pi$ and 
$0 \le \varphi \le 2\pi$, respectively), 
and $f$, $h$ are functions of $r$. 
For the scalar field, we consider a 
radial-dependent profile $\phi(r)$. 
The gravitational and scalar-field equations 
of motion are given by 
\ba
& &
h' = \frac{4(h - 1) [ 1+\kappa
-4h(\phi'' \xi_{,\phi}+\phi'^2 \xi_{,\phi \phi}) ] 
+ hr^2 \phi'^2 + h r [ 4(r \phi'' + 2\phi') \kappa_{,\phi} 
+ 4 r \phi'^2\kappa_{,\phi \phi}]}
{2\left[4 \phi' (3h - 1)\xi_{,\phi}
- r ( r \phi' \kappa_{,\phi}+ 2 \kappa+ 2) \right]}\,,
\label{heq}\\
& &
f' = \frac{f[4(h - 1) (1+\kappa ) 
-h r^2  \phi'^2 + 8 hr \phi' \kappa_{,\phi}]}
{2h\left[4 \phi' (3h - 1)\xi_{,\phi}
- r ( r \phi' \kappa_{,\phi}+ 2 \kappa+ 2) \right]}
\,,\\
& &
\phi''+\left( \frac{2}{r}+\frac{f'}{2f}+\frac{h'}{2h} \right)\phi'
+\frac{R_{{\rm GB}0}^2}{h} \xi_{,\phi}
+\frac{R_{0}}{h} \kappa_{,\phi}=0\,,
\label{phieq}
\ea
where a prime represents the derivative with 
respect to $r$, and
\ba
R_{{\rm GB}0}^2 &=&
\frac{2[(2f f'' - f'^2)h^2 + \{ (3f' h' - 2 f'')f + f'^2 \}h 
- f f' h']}{r^2 f^2}\,,
\label{RGB0} \\
R_{0} &=&
-\frac{(2f f'' - f'^2)h r^2+r f f' (rh' + 4h) 
+4r f^2 h'+4f^2(h - 1)}{2r^2 f^2}\,.
\label{R0} 
\ea
Note that $R_{{\rm GB}0}^2$ and $R_{0}$ are the background 
values of the GB invariant and the Ricci scalar, respectively.
On the Schwarzschild background given by the metric components $f=h=1-r_h/r$, where $r_h$ is the horizon radius, we have $R_{{\rm GB}0}^2=12r_h^2/r^6$ 
and $R_0=0$, respectively. 
If we consider a scalar-field perturbation $\delta \phi$ about the GR branch $\phi=0$, then Eq.~(\ref{phieq0}) gives the linearized equation 
\be
\square \delta \phi+\frac{12 r_h^2}{r^6} \xi_{,\phi \phi} 
\delta \phi=0\,.
\label{delphieq0}
\ee
The GR branch can be subject to tachyonic instability for 
\be
\xi_{,\phi \phi}>0\,.
\ee
The simplest choice of the sGB coupling allowing the existence of 
a nonvanishing $\phi$ branch besides $\phi=0$ is   
$\xi(\phi)=\eta \phi^2/8$. So long as the condition $\eta>0$ 
is satisfied, the GR branch can trigger tachyonic instability 
toward the other branch $\phi \neq 0$. 
This is the phenomenon of spontaneous scalarization of BHs 
induced by the sGB coupling.

In Ref.~\cite{Blazquez-Salcedo:2018jnn}, it was shown that 
the scalarized branch ($\phi \neq 0$) for the coupling 
$\xi(\phi)=\eta \phi^2/8$ is unstable against radial perturbations 
in the even-parity sector, as the tachyonic instability is never 
quenched by nonlinearities in the field equations. 
This can be understood as the appearance of a negative effective 
potential $V_{\rm eff}(r)$ for the scalar-field perturbation 
$\delta \phi$ in the vicinity of the horizon. 
This radial tachyonic instability can be avoided by taking into 
account a coupling proportional to $\phi^4$ in $\xi(\phi)$ \cite{Minamitsuji:2018xde,Silva:2018qhn,Macedo:2019sem}, i.e.,
\be
\xi(\phi)=\frac{\eta}{8} \left( \phi^2+\alpha \phi^4 \right)\,,
\qquad
\kappa(\phi)=0\,,
\label{model1}
\ee
which generates nonlinearies in $\phi$ in the scalar-field equation. 
A negative constant $\alpha$ can lift up the effective potential 
toward the region $V_{\rm eff}(r)>0$. 
Without taking into account the backreaction of metric perturbations 
to the scalar-field perturbation $\delta \phi$, it was shown 
in Ref.~\cite{Minamitsuji:2018xde} that the negative region of 
$V_{\rm eff}(r)$ disappears for $\alpha \phi_0^2 < -0.1155$, 
where $\phi_0$ is the field value on the horizon.
We note that the sGB coupling of the form 
$\xi(\phi)=(\eta/8\beta) (1-e^{-\beta \phi^2})$ can also 
address the radial tachyonic instability problem \cite{Doneva:2017bvd}.
For both the quartic and exponential couplings, since the leading behavior for the small field amplitude is given by $\xi(\phi) \to \eta \phi^2/8$, 
bifurcation of a scalarized branch from the GR Schwarzschild branch occurs 
at the same point in the phase diagram.

There is yet another model of BH spontaneous scalarization based on EsGBR theories given by the coupling functions  \cite{Antoniou:2020nax,Antoniou:2021zoy,Antoniou:2022agj}
\be
\xi(\phi)=\frac{\eta}{8}\phi^2\,,\qquad
\kappa(\phi)=-\frac{\beta}{16}\phi^2\,.
\label{model2}
\ee
In this case, there are also the nontrivial solution $\phi \neq 0$ as well as the GR branch $\phi=0$.
For the latter branch, we have $R_0=0$ and hence the perturbation $\delta \phi$ about $\phi=0$ obeys Eq.~(\ref{delphieq0}). 
Then, under the condition $\eta>0$, the GR solution can 
trigger tachyonic instability toward the other nontrivial branch.
Since the Ricci scalar $R$ acquires the scalar-field contribution for $\phi \neq 0$, the third term on the left-hand side 
of Eq.~(\ref{phieq0}) affects the stability of the branch $\phi \neq 0$. 
Taking into account the backreaction of metric perturbations to the scalar-field perturbation $\delta \phi$, it was shown in Ref.~\cite{Antoniou:2022agj} that\footnote{Our notation of $\beta$ is 4 times as large as that used in Ref.~\cite{Antoniou:2022agj}.} the radially stable BH solutions 
can be present for $\beta>5$. 
Kleihaus {\it et al.} \cite{Kleihaus:2023zzs} found that a new branch of scalarized BHs can emerge from a spherical scalarized branch  for the couplings (\ref{model2}) in EsGBR theories by taking the quadrupolar 
deformation of the horizon configuration (multipole $l=2$) into consideration. 
It is not yet clear whether there are parameter spaces consistent with linear stability conditions along both angular and radial directions, which we will address in the following.

\section{Linear stability conditions for 
high radial and angular momentum modes}
\label{stabilitysec}

To study the linear stability of BHs for the modes with large wavenumbers 
$k$ and multipoles $l$, we consider metric 
perturbations $h_{\mu \nu}$ on the background line element (\ref{spmetric}) given by the metric tensor $\bar{g}_{\mu \nu}$, such that $g_{\mu \nu}=\bar{g}_{\mu \nu}+h_{\mu \nu}$.
We express all the perturbations in terms of the spherical harmonics $Y_{lm} (\theta, \varphi)$, with the parity $(-1)^{l+1}$ for odd modes and the parity $(-1)^l$ for even modes under a rotation in the $(\theta, \varphi)$ plane \cite{Regge:1957td,Zerilli:1970se,Zerilli:1970wzz}. 
Without loss of generality, we set $m=0$ and consider the case in which 
$Y_{l0}$ depends on $\theta$ alone. 
We choose the gauge in which the metric components $h_{ab}$, where the subscripts $a$ and $b$ are either $\theta$ or $\varphi$, are vanishing. 
We also take $h_0=0$, where $h_0$ is the perturbation appearing in the metric components $h_{ta}\,(=h_{at})$ in the form $\sum_{l} h_0 (t,r) \nabla_a Y_{l0} (\theta)$.
These choices completely fix the residual gauge degrees of freedom. 
Then, the nonvanishing metric components are 
\ba
& &
g_{tt}=-f(r)+f(r) \sum_{l} H_0 (t,r) Y_{l0}(\theta)\,,\quad 
g_{tr}=g_{rt}=\sum_{l} H_1 (t,r) Y_{l0}(\theta)\,,\quad
g_{t \varphi}=g_{\varphi t}=- \sum_{l}  Q(t,r) (\sin \theta) 
Y_{l0, \theta} (\theta)\,, \nonumber \\
& &
g_{rr}=h^{-1}(r)+h^{-1}(r)  \sum_{l} H_2(t,r) Y_{l0}(\theta)\,,\quad 
g_{r \theta}=g_{\theta r}= \sum_{l} h_1 (t,r) Y_{l0, \theta}(\theta)\,,
\nonumber \\
& & 
g_{r \varphi}=g_{\varphi r}=-\sum_{l} W(t,r) (\sin \theta) 
Y_{l0, \theta} (\theta)\,, \quad
g_{\theta \theta}=r^2\,,\quad 
g_{\varphi \varphi}=r^2 \sin^2 \theta\,,
\label{gcom}
\ea
where $Q(t,r)$ and $W(t,r)$ correspond to odd-parity perturbations, 
and $H_0(t,r)$, $H_1(t,r)$, $H_2(t,r)$, and $h_1(t,r)$ 
to even-parity perturbations. 
The scalar field contains the even-parity 
perturbation $\delta \phi$, as
\be
\phi=\bar{\phi}(r)+\sum_{l} \delta \phi (t,r) Y_{l0}(\theta)\,,
\ee
where we will omit a bar from the background quantities.

In Horndeski theories given by Eq.~(\ref{action2}), the second-order action containing seven perturbed variables mentioned above was derived in Refs.~\cite{Kobayashi:2012kh,Kobayashi:2014wsa,Kase:2021mix} (see also Refs.~\cite{Kase:2020qvz,Kase:2023kvq} for related works). 
After integrating out some of the nondynamical variables, the resulting quadratic-order action contains one dynamical gravitational perturbation in the odd-parity sector and two dynamical perturbations in the even-parity sector.
This final action, which is expressed in terms of three dynamical 
perturbations, determines conditions for the absence of ghosts 
and Laplacian instabilities in the limits of high radial and angular 
momentum modes.
In the following, we will briefly revisit such conditions and 
apply them to EsGBR theories.

In the odd-parity sector with the multipoles $l \geq 2$, there is one propagating degree of freedom given by 
\be
\chi=\dot{W}-Q'+\frac{2Q}{r}\,,
\ee
where a dot represents the derivative with respect to $t$. 
The dynamical field $\chi$ does not behave as a ghost for 
\ba
{\cal G} &\equiv& 2 
G_4+2 h\phi'^2G_{4,X}-h\phi'^2 \left( G_{5,\phi}
+{\frac {f' h\phi' G_{5,X}}{2f}} \right) 
=1+\kappa-\frac{2f'h
\phi' \xi_{,\phi} }{f} >0\,.
\label{nogoodd1} 
\ea
In the short-wavelength limit, the squared propagation speed 
of $\chi$ along the radial direction is
\be
c_{r,{\rm odd}}^2=\frac{{\cal G}}{{\cal F}}\,,
\label{crodd}
\ee
where the absence of radial Laplacian instability requires that 
\ba
{\cal F} &\equiv& 2 G_4+h\phi'^2G_{5,\phi}-h\phi'^2  
\left( \frac12 h' \phi'+h \phi'' \right) G_{5,X} 
=
1+\kappa-4 \xi_{,\phi \phi}h \phi'^2
-2 (2h \phi''+h' \phi') \xi_{,\phi}
>0\,,
\label{Laprodd} 
\ea
together with the condition (\ref{nogoodd1}). 
In the limit of large multipoles ($l \gg 1$), 
the squared propagation speed of $\chi$ 
along the angular direction is 
\be
c_{\Omega,{\rm odd}}^2
=\frac{{\cal G}}{{\cal H}}\,,
\ee
and hence the angular Laplacian instability is absent for 
\ba
{\cal H} &\equiv& 2 G_4+2 h\phi'^2G_{4,X}-h\phi'^2G_{5,\phi}
-\frac{h^2 \phi'^3 G_{5,X}}{r} 
= 1+\kappa-\frac{4h \phi' \xi_{,\phi}}{r}>0\,,
\label{LapOodd} 
\ea
besides the condition (\ref{nogoodd1}). 

For the monopole ($l=0$), the total second-order action of odd-parity perturbations vanishes identically. 
For the dipole ($l=1$), there is no propagating degree of freedom in the odd-parity sector \cite{Kobayashi:2012kh}.
Hence, in both cases, we do not have additional stability conditions 
besides those derived for $l \geq 2$.

In the even-parity sector with the multipoles $l \geq 2$, there are two dynamical perturbations arising from the matter and gravity sectors. One is the scalar-field perturbation $\delta \phi$, while the other is the gravitational perturbation given by 
\be
\psi=H_2+\frac{a_4}{a_3}l(l+1) h_1+\frac{a_1}{a_3}\delta \phi'\,,
\ee
where 
\be
a_1=\frac{\sqrt{fh}}{2} 
\left[ r^2 \kappa_{,\phi}+4(1-h)\xi_{,\phi} \right]\,,\qquad
a_3=-\frac{1}{2}\phi' a_1-r a_4\,,\qquad 
a_4=\frac{\sqrt{fh}}{2}{\cal H}\,.
\label{a134}
\ee
After integrating out the nondynamical fields $H_0$ and $H_1$ from the second-order action of even-parity perturbations, the resulting quadratic-order action can be expressed in the form 
\be
{\cal S}_{\rm even}=\sum_l\int {\rm d}t\, {\rm d}r 
\left(\dot{\vec{\mathcal{X}}}^{t}{\bm K}\dot{\vec{\mathcal{X}}}
+\vec{\mathcal{X}}'^{t}{\bm G}\vec{\mathcal{X}}'
+\vec{\mathcal{X}}^{t}{\bm S}\vec{\mathcal{X}}'
+\vec{\mathcal{X}}^{t}{\bm M} \vec{\mathcal{X}}\right)\,, 
\label{evenact}
\ee
where $\vec{\mathcal{X}}^{t}=(\psi, \delta \phi)$, and ${\bm K}$, ${\bm G}$, ${\bm M}$ are the $2 \times 2$ symmetric matrices, while ${\bm S}$ is the $2\times 2$ antisymmetric matrix.

The determinants of principal submatrices of ${\bm K}$ are positive for \cite{Kobayashi:2014wsa}
\be
{\cal K} \equiv 2{\cal P}_1-{\cal F}>0\,,
\label{noghosteven}
\ee
where the quantity ${\cal P}_1$ is defined by 
\ba
{\cal P}_1 \equiv \frac{h \mu_s}{2fr^2 {\cal H}^2} \left( 
\frac{fr^4 {\cal H}^4}{\mu_s^2 h} \right)'\,,\qquad 
\mu_s \equiv -\frac{4a_3}{\sqrt{fh}}\,.
\label{defP1P2}
\ea
Under the inequality (\ref{noghosteven}), the ghosts arise 
neither for $\psi$ nor $\delta \phi$. 

In the short-wavelength limit, the radial propagation speeds of $\psi$ and $\delta \phi$ can be obtained by assuming the solutions to the perturbation equations in the form $\vec{\mathcal{X}} \propto \vec{\mathcal{X}}_0 e^{i (\omega t-kr)}$. For large values of $\omega$ and $k$, terms containing the matrix components of ${\bm K}$ and ${\bm G}$ are the dominant contributions to the radial dispersion relation. 
The squared radial propagation speeds of $\psi$ and $\delta \phi$, which are measured in terms of the rescaled radial coordinate $r_s=\int \rd r/\sqrt{h}$ and the proper time $\tau=\int \rd t \sqrt{f}$, are given, respectively, by 
\ba
c_{r1,{\rm even}}^2 &=&
\frac{\cal G}{\cal F}\,,
\label{cr1even}\\
\hspace{-0.5cm}
c_{r2,{\rm even}}^2 &=&
\frac { 4\phi' }{(fh)^{3/2}(2{\cal P}_1-{\cal F}) \mu_s^2}
\left[ 8r^2 h a_4 c_4 (\phi' a_1+ra_4)
-\sqrt{fh}\phi'a_1^2 {\cal G} +2r^2a_4^2 
\left( \frac{f'}{f}a_1+2 c_2\right) \right]\,,
\label{cr2even}
\ea
where 
\be
c_2 =\frac{\sqrt{h}}{4\sqrt{f}} 
\left[ r^2 \phi' f-r (rf'+4f) \kappa_{,\phi}
+4(3h-1)f' \xi_{,\phi} \right]\,,\qquad
c_4 = \frac{1}{2r \sqrt{fh}} 
\left( rf \kappa_{,\phi}-2f' h \xi_{,\phi} 
\right)\,.
\label{c24}
\ee
Since $c_{r1,{\rm even}}^2$ is the same as $c_{r,{\rm odd}}^2$, the Laplacian instability of $\psi$ is absent under the two conditions (\ref{nogoodd1}) and (\ref{Laprodd}). 
The radial Laplacian stability of $\delta \phi$ requires that 
\be
c_{r2,{\rm even}}^2>0\,.
\ee

In the limit of large multipoles $l$ and high frequencies $\omega$, the dominant contributions to the angular dispersion relation arise from the matrix components of ${\bm K}$ and ${\bm M}$.
The squared angular propagation speeds of even-parity dynamical perturbations for $l \gg 1$, which are measured in terms of the infinitesimal angular distance $r \rd \theta$ and the proper time $\rd \tau$, are expressed, respectively, by 
\be
c_{\Omega \pm}^2=-B_1\pm\sqrt{B_1^2-B_2}\,, 
\label{cosqeven}
\ee
where 
\ba
\hspace{-0.8cm}
&&
B_1=
{\frac { a_4 r^3 [ 4 h ( \phi' a_1+2 ra_4 ) \beta_1+\beta_2
-4 \phi' a_1 \beta_3  ] 
-2 fh{\cal G}  [ 2 ra_4  ( 2 {\cal P}_1-{\cal F} )  ( \phi' a_1+ra_4 ) 
+\phi'^2a_1^{2}{\cal P}_1 ] }
{ 4\sqrt {fh} a_4( \phi' a_1+2 ra_4 )^2 ( 2 {\cal P}_1-{\cal F} )}},
\label{B1def}\\
&&
B_2=
-r^2{\frac {r^2h \beta_1 [ 2 fh {\cal F} {\cal G} ( \phi' a_1+2 ra_4 ) +r^2\beta_2 ] 
-{r}^{4}\beta_2 \beta_3
-fh{\cal F} {\cal G}  ( \phi' fh {\cal F} {\cal G}a_1 +4 r^3 a_4 \beta_3 ) }
{ fh{\cal F} \phi' a_1  ( \phi' a_1+2 ra_4 ) ^{2} ( 2 {\cal P}_1-{\cal F} )  }}\,,
\label{B2def}
\ea
where $\beta_1$, $\beta_2$, $\beta_3$ are given in Appendix A. 
To ensure the angular Laplacian stability, we require that $c_{\Omega+}^2>0$ and $c_{\Omega-}^2>0$.  
These conditions are satisfied if
\be
B_1^2 \geq B_2>0 \quad {\rm and} \quad 
B_1<0\,.
\label{B12con}
\ee
The violation of the condition $B_1^2-B_2 \geq 0$ gives rise to the imaginary values of $c_{\Omega \pm}^2$, one of which leads to the Laplacian instability. 
Even for $B_1^2-B_2 \geq 0$, the inequalities $B_1<0$ and $B_2>0$ need to be satisfied further to ensure that 
$c_{\Omega \pm}^2>0$. 

Before ending this section, we consider theories given by the coupling functions 
\be
\xi(\phi)=\frac{\eta}{8} \left( \phi^2+\alpha \phi^4 \right)\,,
\qquad
\kappa(\phi)=-\frac{\beta}{16}\phi^2\,,
\label{model12}
\ee
which accommodate both the models (\ref{model1}) and (\ref{model2}). 
For the scalarized BH solution where $|\phi(r)|$ is much smaller than 1, we expand the quantities ${\cal G}$, ${\cal F}$, ${\cal H}$, ${\cal K}$, and $c_{r2,{\rm even}}^2$ around $\phi=0$.
On using the background Eqs.~(\ref{heq})-(\ref{phieq}) to eliminate the derivative terms $f'$, $h'$, and $\phi''$, it follows that 
\ba
{\cal G} &=& 1-\frac{\eta (4-4h
+h r^2\phi'^2)\phi'}{8r}\phi+{\cal O}(\phi^2)\,,\\
{\cal F} &=& 1-\eta h \phi'^2
+\frac{\eta (4+12h
+h r^2\phi'^2)\phi'}{8r}\phi+{\cal O}(\phi^2)\,,\\
{\cal H} &=& 1-\frac{\eta h\phi'}{r}\phi+{\cal O}(\phi^2)\,,\\
{\cal K} &=& \frac{r^2 \phi'^2}{4} \left[ 
1+\frac{\{8\eta (h-1)+\beta r^2 \}\phi'}{8r}\phi
+{\cal O}(\phi^2)
\right]\,,\label{Kexpan}\\
c_{r2,{\rm even}}^2 &=& 1-\frac{\eta h \phi'^2}{64} 
\left[ \beta+\frac{8 \eta (h-1)}{r^2} \right]
\left[ 3\beta-8+\frac{24 \eta (h-1)}{r^2} \right]\phi^2
+{\cal O}(\phi^3)\,.
\label{crap}
\ea
We study the case in which $\phi (r)$ is a positive decreasing function of $r$ outside the horizon ($r \ge r_h$), with $|r\phi'|$ being at most of order 
$\phi$. Then, the scalarized solution has the largest field value $\phi_0$ 
on the horizon ($r=r_h$).  
So long as the condition
\be
\bar{\eta} \phi_0^2 \ll 1
\label{phicon1}
\ee
is satisfied, where
\be
\label{bareta}
\bar{\eta} \equiv \frac{\eta}{r_h^2}
\ee
is a dimensionless coupling constant, all of the dominant contributions to ${\cal G}$, ${\cal F}$, and ${\cal H}$ are 1 outside the horizon. 
If the inequality
\be
\beta \phi_0^2 \ll 1
\label{phicon2}
\ee
holds in addition to the condition (\ref{phicon1}), the dominant contribution to ${\cal K}$ is $r^2 \phi'^2/4$ and hence it is positive. 
If the two inequalities (\ref{phicon1}) and (\ref{phicon2}) are satisfied, then $c_{r2,{\rm even}}^2$ is close to 1 for 
\be
\bar{\eta} \lesssim {\cal O}(1)\,.
\label{etacon}
\ee
Thus, the stability conditions ${\cal G}>0$, ${\cal F}>0$, ${\cal H}>0$, ${\cal K}>0$, and $c_{r2,{\rm even}}^2>0$ hold for the field value $\phi_0$ and the couplings $\bar{\eta}$, $\beta$ in the ranges (\ref{phicon1}), (\ref{phicon2}), and (\ref{etacon}).
Since the angular stability conditions associated with $c_{\Omega \pm}^2$ are more involved, we will study them in detail in Secs.~\ref{angsec1} and \ref{angsec2}.

\section{Radial tachyonic stability of scalarized BHs}
\label{radisec}

In this section, we will revisit the radial tachyonic stability of scalarized BHs in both EsGB and EsGBR theories. 
This amounts to studying a monopole perturbation without restricting the analysis to the short-wavelength limit $k r_h \gg 1$. In other words, we discuss the radial stability of the dynamical perturbation $\delta \phi$ for  $l=0$ by paying particular attention to an effective potential induced by its effective mass term. For $l=0$, it was shown in Refs.~\cite{Kobayashi:2014wsa,Kase:2021mix,Kase:2023kvq} that the gravitational perturbation $\psi$ does not propagate and hence the scalar-field perturbation $\delta \phi$ is the only propagating  degree of freedom in the even-parity sector\footnote{For the dipole mode $l=1$, we can choose the gauge $\delta \phi=0$ to fix the residual gauge degree of freedom. 
In this case, $\psi$ is the only propagating degree of freedom in the even-parity sector.
Again, the no-ghost condition and the radial propagation speed of even-parity perturbations 
are the same as Eqs.~(\ref{noghosteven}) and (\ref{cr2even}), respectively \cite{Kobayashi:2014wsa,Kase:2021mix,Kase:2023kvq}.}. 
As we will see below, the radial propagation speed squared $c_r^2$ for $l=0$ 
is equivalent to Eq.~(\ref{cr2even}) obtained for $l \geq 2$.

In Horndeski theories, the second-order action of even-parity perturbations was derived in Refs.~\cite{Kobayashi:2014wsa,Kase:2021mix,Kase:2023kvq} for the perturbed metric components (\ref{gcom}). 
Focusing on the monopole mode ($l=0$), the action is expressed in the form ${\cal S}_{\rm even}=\int \rd t \rd r\,{\cal L}$, where
\ba
{\cal L} &=&
H_0 \left[ a_1 \delta \phi'' +a_2 \delta \phi'
+ (a_2'-a_1'') \delta \phi+a_3 H_2'+a_3' H_2 \right]
-\frac{2}{f} H_1 \left[ a_1 \dot{\delta \phi}'
+(a_2-a_1') \dot{\delta \phi}+a_3 \dot{H}_2 \right]
\nonumber \\
& &+c_1 \dot{\delta \phi} \dot{H}_2+
\left( c_2 \delta \phi'+c_3 \delta \phi \right)H_2
+c_6 H_2^2+e_1 \dot{\delta \phi}^2+e_2 \delta \phi'^2
+e_3 \delta \phi^2\,,
\label{Lag}
\ea
where the explicit forms of the $r$-dependent coefficients are given in Eqs.~(\ref{a134}) and (\ref{c24}), and Eqs. (\ref{coeff1}) and (\ref{coeff2}) in Appendix A. For $l=0$, the contributions to ${\cal L}$ arising from the metric 
perturbation $h_1$ identically vanish. 
Hence we have the three metric perturbations $H_0$, $H_1$, $H_2$, and the scalar-field perturbation $\delta \phi$ in the second-order action of even-parity perturbations.  
We note that the radial stability of scalarized BHs was also discussed in Refs.~\cite{Blazquez-Salcedo:2018jnn,Silva:2018qhn,Antoniou:2021zoy,Antoniou:2022agj} 
by considering the perturbations $H_0$, $H_2$, and $\delta \phi$. 
While we need to include the field $H_1$ for the consistent analysis,  it does not eventually contribute to the dynamics of perturbations as we will see below.

Let us introduce the following perturbed quantity
\be
\Phi=a_1 \delta \phi'+\left( a_2-a_1' \right) 
\delta \phi+a_3 H_2\,.
\ee
Then, the Lagrangian density (\ref{Lag}) can be written as 
\be
{\cal L}=
H_0 \Phi'-\frac{2}{f}H_1 \dot{\Phi}
+c_1 \dot{\delta \phi} \dot{H}_2
+\left( c_2 \delta \phi'+c_3 \delta \phi \right)H_2
+c_6 H_2^2+e_1 \dot{\delta \phi}^2+e_2 \delta \phi'^2
+e_3 \delta \phi^2\,.
\label{Lag1}
\ee
Varying ${\cal L}$ with respect to $H_0$ and $H_1$, 
respectively, we obtain
\be
\Phi'=0\,,\qquad \dot{\Phi}=0\,,
\ee
so that $\Phi={\cal C}={\rm constant}$. 
Since the integration constant ${\cal C}$ is irrelevant to the perturbation dynamics, we set ${\cal C}=0$ in the following. Then, we have
\be
a_1 \delta \phi'+\left( a_2-a_1' \right) 
\delta \phi+a_3 H_2=0\,,
\label{phiH2}
\ee
with the Lagrangian density  
\be
{\cal L}=c_1 \dot{\delta \phi} \dot{H}_2
+\left( c_2 \delta \phi'+c_3 \delta \phi \right)H_2
+c_6 H_2^2+e_1 \dot{\delta \phi}^2+e_2 \delta \phi'^2
+e_3 \delta \phi^2\,.
\label{Lag2}
\ee
Solving Eq.~(\ref{phiH2}) for $H_2$ and taking its time derivative, we can eliminate the $H_2$-dependent terms in Eq.~(\ref{Lag2}) to give 
\be
{\cal L}=K_0 \dot{\delta \phi}^2-G_0 \delta \phi'^2
-M_0 \delta \phi^2+S_0 \delta \phi' \delta \phi\,,
\label{Lag3}
\ee
where 
\ba
K_0 &=& e_1+\frac{[ a_1 c_1'+(3a_1'-2a_2) c_1] a_3 
-a_1 c_1 a_3'}{2a_3^2}\,,
\label{K0}\\
G_0 &=& -e_2+\frac{a_1 (a_3 c_2-a_1 c_6)}{a_3^2}\,,\\
M_0 &=& -e_3+\frac{(a_2 - a_1')[c_3 a_3-c_6(a_2 - a_1')]}{a_3^2}\,,
\label{M0} \\
S_0 &=& \frac{(a_2-a_1')(2a_1 c_6-a_3 c_2)-a_1 a_3 c_3}{a_3^2}\,.
\label{S0}
\ea
Thus, for $l=0$, the dynamics of even-parity perturbations is 
governed by the single propagating  degree of freedom $\delta \phi$. 
The ghost is absent under the condition $K_0>0$. 
Since $K_0$ can be expressed as $K_0=(2{\cal P}_1-{\cal F})/(\sqrt{fh}\,\phi'^2)$ \cite{Kase:2023kvq}, the no-ghost condition translates to 
$2{\cal P}_1-{\cal F}>0$. 
This is equivalent to the condition (\ref{noghosteven}) derived for $l \geq 2$.
The radial propagation speed squared of $\delta \phi$ measured in terms of the rescaled radial coordinate $r_s=\int \rd r/\sqrt{h}$ and the proper time 
$\tau=\int \rd t \sqrt{f}$ is given by 
\be
c_r^2=\frac{G_0}{fh K_0}\,,
\ee
which is equivalent to $c_{r2,{\rm even}}^2$ obtained for $l \geq 2$. 
Thus the radial Laplacian instability is absent for $c_r^2>0$, which translates to $G_0>0$ under the no-ghost  condition $K_0>0$. 

Varying (\ref{Lag3}) with respect to $\delta \phi$, it follows that 
\be
g_s^2(r) \ddot{\delta \phi}-\delta \phi''+{\cal C}_1(r) 
\delta \phi'+U(r) \delta \phi=0\,,
\label{delphieq}
\ee
where 
\be
g_s^2 (r) \equiv \frac{K_0}{G_0}=\frac{1}{fh\,c_r^2}\,,\qquad 
{\cal C}_1(r) \equiv -\frac{G_0'}{G_0}\,,\qquad 
U(r) \equiv \frac{2M_0+S_0'}{2G_0}\,.
\label{Udef}
\ee
To study the radial stability of $\delta \phi$ associated with the potential $U(r)$, we consider the solution to  Eq.~(\ref{delphieq}) in the form 
\be
\delta \phi (t,r)={\cal C}_0(r) \delta \varphi (r) e^{-i \omega t}\,,
\ee
where $\omega$ is an angular frequency, and ${\cal C}_0(r)$ and $\delta \varphi(r)$ are $r$-dependent functions. 
We introduce the tortoise coordinate $r_*=\int g_s\,{\rm d}r$ and choose ${\cal C}_0(r)$ to eliminate the first derivative $\rd \delta \varphi/\rd r_*$ in the equation of motion for $\delta \varphi$ following from Eq.~(\ref{delphieq}). 
Then, we obtain
\be
\frac{{\cal C}_0'}{{\cal C}_0}=\frac{1}{2} 
\left( {\cal C}_1-\frac{g_s'}{g_s} \right)\,, 
\ee
and 
\be
\left( -\frac{\rd^2}{\rd r_*^2}+V_{\rm eff} 
\right) \delta \varphi=\omega^2 \delta \varphi\,,
\ee
where 
\be
V_{\rm eff}=\frac{1}{g_s^2} \left( U+\frac{{\cal C}_1^2}{4}
-\frac{{\cal C}_1'}{2}-\frac{3g_s'^2}{4g_s^2}
+\frac{g_s''}{2g_s} \right)\,.
\label{Veff}
\ee
If the effective potential $V_{\rm eff}(r_*)$ is negative at some distance $r_*$, it signals the presence of radial tachyonic instability.
In the following, we will study the shapes of $V_{\rm eff}(r)$ in both EsGB theories and EsGBR theories.

\subsection{EsGB theories}
\label{EsGBrasec}

In EsGB theories charactrized by the coupling functions 
$\xi(\phi) \neq 0$ and $\kappa(\phi)=0$, the quantity 
$e_3$ appearing in the mass term $M_0$ in Eq.~(\ref{M0}) 
is given by 
\be
e_3= \frac{r^2 \sqrt{f}}{4 \sqrt{h}} 
\xi_{,\phi \phi} R_{{\rm GB}0}^2\,.
\ee
The dominant contributions to $f$ and $h$ are the Schwarzschild metric components $f=h=1-r_h/r$.
Then, the leading-order contribution to the background 
GB invariant (\ref{RGB0}) is given by $R_{{\rm GB}0}^2=12r_h^2/r^6$. 
The metric perturbation $H_2$ affects the 
quantities $K_0$, $G_0$, and $M_0$ through the 
terms containing $c_1$, $c_2$, $c_3$, and $c_6$ in 
Eqs.~(\ref{K0})-(\ref{S0}).
If we neglect the backreaction of metric perturbations 
to $\delta \phi$, the potential $U$ in Eq.~(\ref{Udef}) 
is approximately given by 
\be
U \simeq \frac{M_0}{G_0} \simeq \frac{e_3}{e_2}
=-\frac{1}{h}\xi_{,\phi \phi} R_{{\rm GB}0}^2\,,
\ee
where we used $e_2=-r^2 \sqrt{fh}/4$.

Let us consider the coupling $\xi(\phi)=(\eta/8)(\phi^2+\alpha \phi^4)$ 
and focus on the regime in which the two conditions 
(\ref{phicon1}) and (\ref{etacon}) hold, i.e., 
\be
\bar{\eta} \phi_0^2 \ll 1\,,\qquad 
\bar{\eta} \lesssim {\cal O}(1)\,,
\label{phi0ra}
\ee
where we recall that $\phi_0$ is the field value on the horizon. 
Then, as we showed in Eq.~(\ref{crap}), $c_{r}^2$ is close to 1. 
In this regime, we can exploit the approximation that the backreaction 
of the scalar field on the metric is negligible. 
This means that, for the estimations of $g_s$ and ${\cal C}_1$, 
the metric components $f$ and $h$ 
are approximated as $f \simeq h \simeq 1-r_h/r$. 
Since we have $g_s \simeq (1-r_h/r)^{-1}$ and 
$G_0 \simeq (r^2/4)(1-r_h/r)$, 
the effective potential (\ref{Veff}) reduces to 
\be
V_{\rm eff} \simeq \frac{r_h (r-r_h)}{r^7} 
(r^3-12 r_h \xi_{,\phi \phi})\,,
\label{Veffap}
\ee
where $\xi_{,\phi \phi}=\eta(1+6 \alpha \phi^2)/4$.
This approximate effective potential coincides with 
the one derived in Ref.~\cite{Minamitsuji:2018xde}. 
For $\alpha=0$ and $\bar{\eta}={\cal O}(1)$, we have 
$V_{\rm eff} (r)<0$ 
in the vicinity of the horizon and hence the radial instability 
is present for the scalarized branch. 
For $\alpha<0$, it is possible to realize 
$V_{\rm eff}(r)>0$ throughout the horizon exterior. 
The analysis of Ref.~\cite{Minamitsuji:2018xde} based on the 
approximate potential (\ref{Veffap}) shows that the negative 
region of $V_{\rm eff}(r)$ disappears for 
\be
\alpha \phi_0^2 < -0.1155\,.
\label{alcon}
\ee
This condition was derived by neglecting the backreaction of metrics 
to the scalar-field perturbation $\delta \phi$.

\begin{figure}[ht]
\begin{center}
\includegraphics[height=3.4in,width=3.4in]{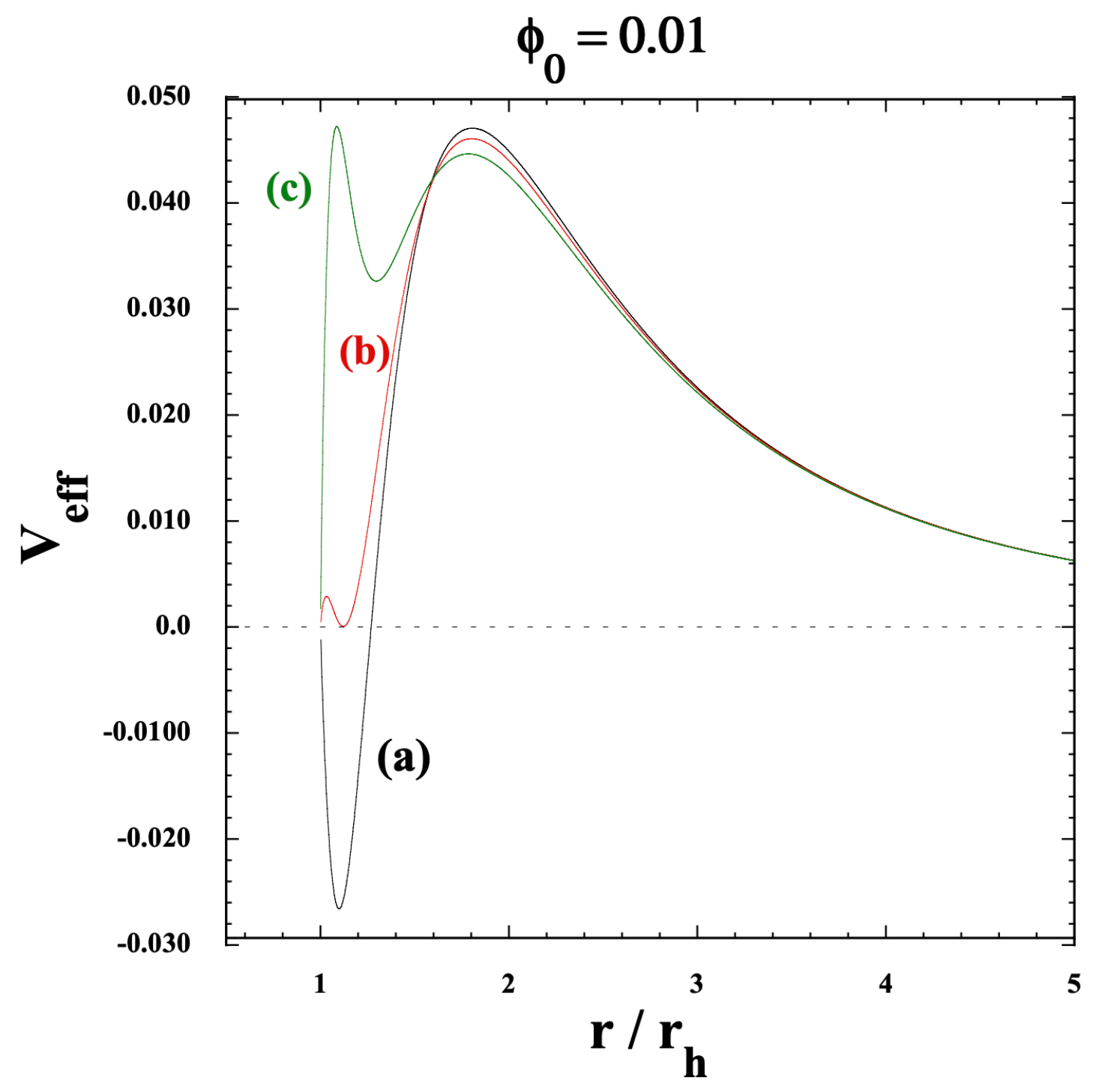}
\end{center}
\caption{\label{VeffEsGB} 
The effective potential $V_{\rm eff}$ versus $r/r_h$ 
in EsGB theories with the coupling  
$\xi(\phi)=(\eta/8)(\phi^2+\alpha \phi^4)$ and 
the field value $\phi_0=0.01$ on the horizon.
Each case corresponds to the model parameters
(a) $\bar{\eta}=0.7742$, $\alpha=-500$, 
(b) $\bar{\eta}=0.8500$, $\alpha=-1155$, and 
(c) $\bar{\eta}=0.9807$, $\alpha=-2000$, respectively. 
For $\alpha \phi_0^2 < -0.1155$, the effective potential 
is positive throughout the horizon exterior.
}
\end{figure}

In order to confirm whether the condition (\ref{alcon}) gives 
a good criterion for the radial tachyonic stability of $\delta \phi$,
we numerically compute $V_{\rm eff} (r)$ in Eq.~(\ref{Veff}) 
without using the approximations for $U$, ${\cal C}_1$, and $g_s$
explained above.
In Fig.~\ref{VeffEsGB}, we plot $V_{\rm eff}$ versus $r/r_h$ 
for $\phi_0=0.01$ by choosing three different combinations 
of $(\bar{\eta}, \alpha)$. 
In case (a), where the condition (\ref{alcon}) is violated, 
we have $V_{\rm eff}(r)<0$ in the vicinity of $r=r_h$. 
Case (b) corresponds to the border value 
$\alpha \phi_0^2=-0.1155$, below which the region 
with $V_{\rm eff}(r)<0$ starts to disappear.
In case (c), where the inequality (\ref{alcon}) is satisfied,  
$V_{\rm eff}(r)$ is positive for $r \geq r_h$.
Thus, the full analysis of the effective potential shows 
that the condition (\ref{alcon}) is a trustable criterion 
for the radial stability of scalarized BHs.
We confirm that this is also the case for other 
field values $\phi_0$ in the range (\ref{phi0ra}).

\subsection{EsGBR theories}
\label{EsGBRbo}

In EsGBR theories given by the coupling functions $\xi(\phi)=\eta \phi^2/8$ and $\kappa (\phi)=-\beta \phi^2/16$, the quantity $e_3$ in $M_0$ is given by 
\be
e_3=\frac{r^2 \sqrt{f}}{32\sqrt{h}} 
\left( 2\eta R_{{\rm GB}0}^2-\beta R_0 \right)\,.
\label{e2}
\ee
On the GR branch ($\phi=0$) with the Schwarzscild background ($f=h=1-r_h/r$), we have $R_{{\rm GB}0}^2=12r_h^2/r^6$ and $R_0=0$. 
Then, for $\eta>0$, this GR solution can be subject to tachyonic instability toward the scalarized branch ($\phi \neq 0$) due to the dominance of a negative term in $M_0$ induced by the sGB coupling.
After the solution reaches the scalarized branch, the Ricci scalar $R_0$ acquires the contribution of $\phi$ through the coupling between gravity and the scalar field. 
For $\beta>0$, the $-\beta R_0$ term in 
$e_3$ gives rise to a positive contribution to $M_0$. 

Moreover, since we are considering the scalar-field contribution to the gravity sector at the background 
level, we cannot generally neglect the backreaction of the metric perturbation $H_2$ on the dynamics of $\delta \phi$. In other words, we should exploit 
the full expressions of $K_0$, $G_0$, $M_0$, and $S_0$ in Eqs.~(\ref{K0})-(\ref{S0}) to accurately compute the quantities $U$, ${\cal C}_1$, and $g_s^2$ appearing in the effective potential $V_{\rm eff}(r)$. 
Indeed, unlike EsGB theories discussed in Sec.~\ref{EsGBrasec}, the analysis 
based on ignoring the backreaction of $H_2$ on $\delta \phi$ leads to different forms of $V_{\rm eff}(r)$ especially in the vicinity of the horizon.
In other words, even if the full analysis without using any approximation gives $V_{\rm eff}(r)>0$ throughout the horizon exterior, it can happen that the approximation neglecting the metric backreaction generates the region of negative values of $V_{\rm eff}(r)$.

\begin{figure}[ht]
\begin{center}
\includegraphics[height=3.4in,width=3.4in]{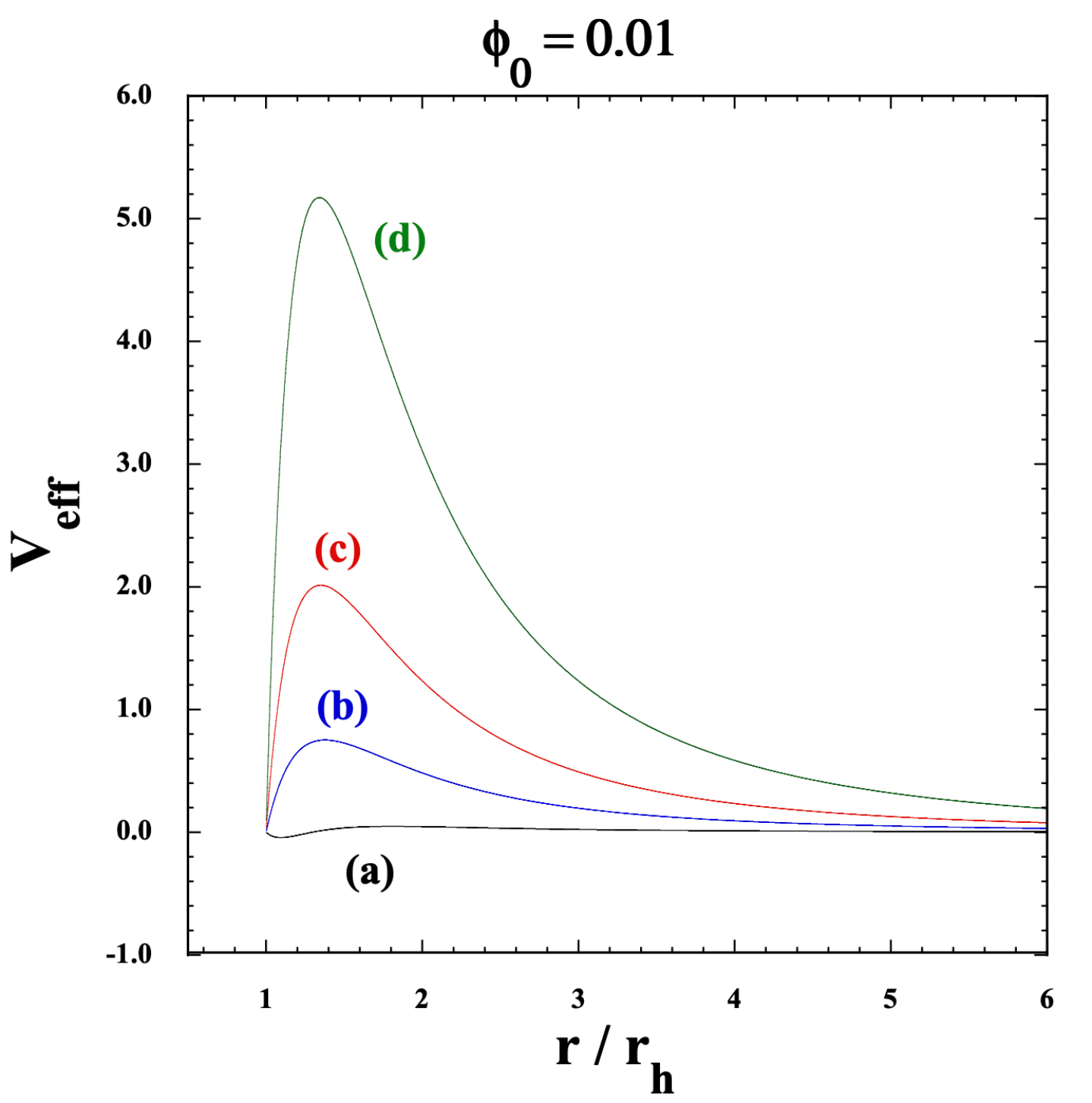}
\end{center}
\caption{\label{VeffEsGBR} 
The effective potential $V_{\rm eff}$ versus $r/r_h$ 
in EsGBR theories with the couplings  
$\xi(\phi)=\eta\phi^2/8$ and $\kappa(\phi)=-\beta \phi^2/16$ 
for the field value $\phi_0=0.01$ on the horizon.
Each case corresponds to the model parameters
(a) $\bar{\eta}=0.72565$, $\beta=1$, 
(b) $\bar{\eta}=0.72568$, $\beta=8$, 
(c) $\bar{\eta}=0.72612$, $\beta=20$, and 
(d) $\bar{\eta}=0.72701$, $\beta=50$ 
respectively. }
\end{figure}

In Fig.~\ref{VeffEsGBR}, we plot $V_{\rm eff}$ versus $r/r_h$ for $\phi_0=0.01$ with four different combinations of $(\bar{\eta}, \beta)$. 
Note that this effective potential is numerically computed without 
neglecting the backreaction of $H_2$ on $\delta \phi$. 
For $\beta=1$, there is the region with $V_{\rm eff}(r)<0$ around the horizon. In other three cases, which correspond to $\beta=8, 20, 50$, 
we have $V_{\rm eff}(r)>0$ throughout the horizon exterior. 
This means that, with the field value $\phi_0=0.01$, the radial tachyonic instability of $\delta \phi$ is absent for $\beta>{\cal O}(1)$. 
The study of Ref.~\cite{Antoniou:2022agj} shows that 
the radial stability is ensured for
\be
\beta>5\,,
\label{betara}
\ee
irrespective of the field value $\phi_0$ on the horizon. 
Our analysis is consistent with this condition. 
Even with small values of $|\phi_0|$ of order $10^{-3}$, 
we confirm that $V_{\rm eff}(r)>0$ outside the horizon 
for $\beta$ in the range (\ref{betara}). 
We stress that this positivity of $V_{\rm eff}(r)$ in the range $|\phi_0| \ll 1$ arises by consistently incorporating the backreaction of metric 
perturbations on the dynamics of $\delta \phi$. 
Interestingly, the absence of radial tachyonic instability of $\delta \phi$ does not particularly impose the minimal values of $|\phi_0|$. This property is different from EsGB theories discussed in Sec.~\ref{EsGBrasec}, where the amplitude of $\phi_0$ is constrained to be $|\phi_0|>\sqrt{0.1155/|\alpha|}$.

\section{BH linear stability in EsGB theories}
\label{angsec1}

In this section, we investigate the angular Laplacian stability 
of scalarized BHs in EsGB theories given by the couplings (\ref{model1}). 
We first analytically study the Laplacian stability conditions far away from the horizon ($r \gg r_h$). 
Then, we consider the regime in the vicinity of the horizon and find the parameter region in which both 
$c_{\Omega +}^2$ and $c_{\Omega -}^2$
 are positive. We also numerically compute 
$c_{\Omega \pm}^2$ outside the horizon and study their behavior in an intermediate regime. 
Then, we elucidate the parameter space in which the 
scalarized BHs are linearly stable in both angular 
and radial directions.

\subsection{Region far away from the horizon}
\label{infexp}

At spatial infinity, we impose a boundary condition of the vanishing scalar field, i.e., $\phi (\infty)=0$. 
At large distances the scalar field acquires a charge $q_s$, so that the solution can be expressed in the form $\phi=q_s/r+{\cal O}(r^{-2})$. The Schwarzschild metric components $f=h=1-m/r$, where $m$ corresponds to 
twice the Arnowitt-Deser-Misner (ADM) mass, are modified by 
the scalar charge (note that $m=r_h$ for the Schwarzschild metric). 
The background solutions to $\phi$, $f$, and $h$, 
which are expanded up to the fifth order in $1/r$, are given by 
\ba
\phi &=& \frac{q_s}{r}+\frac{q_s m}{2r^2}
-\frac{q_s (q_s^2-8 m^2
)}{24r^3}-\frac{q_s m
(q_s^2-3m^2)}{12r^4}
+\frac{9q_s^5-8q_s(29q_s^2+36 \eta) m^2 
+384 q_s m^4}
{1920\,r^5}+{\cal O} (r^{-6})\,,\\
f &=& 1-\frac{m}{r}+\frac{q_s^2 m}{24r^3}
+\frac{q_s^2 m^2}{24r^4}
-\frac{q_s^2 m (3q_s^2 -24 m^2-256 \eta)}{640 r^5}
+{\cal O} (r^{-6})\,,\\
h &=& 1-\frac{m}{r}+\frac{q_s^2}{4r^2}
+\frac{q_s^2 m}{8r^3}+\frac{q_s^2 m^2}{12r^4}
-\frac{q_s^2 m (q_s^2-24m^2
-384 \eta)}{384 r^5}
+{\cal O} (r^{-6})\,.
\ea
While the coupling constant $\eta$ appears at the fifth-order expansions 
of $\phi$, $f$, $h$ in $1/r$, the other coupling constant $\alpha$ arises 
at their seventh-order expansions. 
In order to correctly estimate the quantities associated with stability of scalarized BHs, we need to expand $\phi$, $f$, $h$ 
at least up to ninth order, which will be used in the following.

The quantities associated with the stability of odd-parity perturbations are given by 
\be
{\cal G}=1+\frac{\eta q_s^2 m}{2r^5}+{\cal O}(r^{-6})\,,\qquad 
{\cal F}=1-\frac{3\eta q_s^2}{r^4}+{\cal O}(r^{-5})\,,\qquad 
{\cal H}=1+\frac{\eta q_s^2}{r^4}+{\cal O}(r^{-5})\,,
\ee
whose leading-order terms are all positive. 

In the even-parity sector, the no-ghost quantity (\ref{noghosteven}) yields 
\be
{\cal K}=\frac{q_s^2}{4r^2}+\frac{q_s^2 m}{2r^3}
+{\cal O}(r^{-4})\,,
\label{calK}
\ee
so that ${\cal K}>0$ at leading order. 
The radial propagation speed squared (\ref{cr2even}) is estimated as 
$c_{r2,{\rm even}}^2=1+{\cal O}(r^{-8})$ and hence it is very close to 1.
The quantities $B_1$ and $B_2$ defined in Eqs.~(\ref{B1def}) and (\ref{B2def}) are given, respectively, by 
\ba
B_1 &=& -1+\frac{\eta q_s^2}{2r^4}
-\frac{\eta [5q_s^2 m^2
-(1+24\alpha)q_s^4]}{24r^6}
+\frac{\eta q_s^2 m
[(1+12\alpha)q_s^2-4m^2]}{12 r^7} 
\nonumber \\
& &
-\frac{\eta q_s^2 [q_s^4+3q_s^2 
\{40 \eta-(11+40\alpha) m^2
\}
+101 m^4+1188 \eta m^2]}{240 r^8}+{\cal O}(r^{-9})\,,\label{B1a}\\
B_2 &=& 1-\frac{\eta q_s^2}{r^4}+\frac{\eta [5q_s^2 m^2
-(1+24\alpha)q_s^4]}{12r^6}
-\frac{\eta q_s^2 m [(1+12\alpha)q_s^2-4m^2]}{6 r^7} 
\nonumber \\
& &
+\frac{\eta q_s^2 [q_s^4+3q_s^2 \{40 \eta-(11+40\alpha)
m^2\}+101m^4+108 \eta m^2
]}{120 r^8}+{\cal O}(r^{-9})\,.
\ea
Then, it follows that 
\be
B_1^2-B_2=\frac{\eta^2 q_s^2 (q_s^2+36 m^2)}
{4r^8}+{\cal O}(r^{-9})\,,
\label{B12a}
\ee
whose leading-order contribution to $B_1^2-B_2$ is positive\footnote{There was an error in the calculation of Ref.~\cite{Minamitsuji:2023uyb}, which led to the incorrect result that the leading-order term of $B_1^2-B_2$ is negative. In deriving the results in Ref.~\cite{Minamitsuji:2023uyb}, there was a typo in the input of the coefficient $d_3$ defined in Eq.~\eqref{d3} in the Mathematica code. This led to a wrong result in the ${\cal O} (r^{-7})$ terms of $B_1$ and $B_2$. After correcting this, the ${\cal O}(r^{-7})$ terms in $B_1^2-B_2$ cancel out, which gives rise to no instabilities in the angular directions in the large-distance limit.}.
Our new result (\ref{B12a}) means that both $c_{\Omega+}^2$ and $c_{\Omega-}^2$ are real. Since $B_1$ and $B_2$ approach $-1$ and $1$ as $r \to \infty$, respectively, the conditions (\ref{B12con}) are all satisfied at large distances and hence there are no angular Laplacian instabilities of even-parity perturbations. 

\subsection{Region in the vicinity of the horizon}
\label{hoexp}

Around the horizon characterized by the distance $r_h$, we expand the metric components and the scalar field, as 
\be
f=\sum_{i=1} f_i (r-r_h)^i\,,\qquad 
h=\sum_{i=1} h_i (r-r_h)^i\,,\qquad 
\phi=\sum_{i=0} \phi_i (r-r_h)^i\,,
\label{fhho}
\ee
where $f_i$, $h_i$, and $\phi_i$ are constants. 
The first component $f_1$ is not particularly constrained from the background equations, while $h_1$ and $\phi_1$ are given by 
\be
h_1=-\frac{\phi_1}{3 \bar{\eta} \phi_0 (2\alpha \phi_0^2+1)}\,,
\qquad 
\phi_1=-\frac{1-\sqrt{1-6\bar{\eta}^2 \phi_0^2(2\alpha \phi_0^2+1)^2}}
{\bar{\eta}r_h \phi_0 (2\alpha \phi_0^2+1)}\,,
\ee
where $\bar{\eta}$
is defined in Eq.~\eqref{bareta}.
To realize a scalar-field profile where $|\phi(r)|$ is a decreasing function of $r$ around the horizon, we require that $\phi_0 \phi_1<0$.
For $\bar{\eta}>0$, this demands the inequality $2\alpha \phi_0^2+1>0$. 
Combining this with Eq.~(\ref{alcon}), we obtain
\be
-\frac{1}{2}<\alpha \phi_0^2<-0.1155\,.
\label{alphi}
\ee
whose conditions are required for the existence of hairy BH solutions without the radial tachyonic instability. 
To ensure that $\phi_1$ is real, we also need the inequality $6\bar{\eta}^2 \phi_0^2(2\alpha \phi_0^2+1)^2<1$. 
For $\bar{\eta} \phi_0$ smaller than the order 1, this condition is satisfied under the bound (\ref{alphi}).

On using the expansion (\ref{fhho}), we can estimate ${\cal G}$, ${\cal F}$, and ${\cal H}$ at the horizon $r=r_h$ to be
\be
{\cal G}(r_h)={\cal F}(r_h)=\frac{2}
{\sqrt{1-6 \bar{\eta}^2 \phi_0^2 
(2\alpha \phi_0^2+1)^2}+1}\,,\qquad 
{\cal H}(r_h)=1\,,
\ee
which are positive. 
In the regime where the inequality $\bar{\eta}\phi_0^2 \ll 1$ is satisfied, the leading-order contribution to Eq.~(\ref{Kexpan}) on the horizon is given by 
\be
{\cal K}(r_h)=\frac{r^2 \phi(r)'^2}{4} \biggr|_{r=r_h}=
\frac{9 \bar{\eta}^2 \phi_0^2}{4}\,,
\label{Krh}
\ee
which is positive. 
We also find that $c_{r2,{\rm even}}^2$ is equivalent to 1 at $r=r_h$.
As for the squared angular propagation speeds in the even-parity sector on the horizon, the expansions of $B_1$ and $B_2$ for small values of $\phi_0$ close to 0 give 
\be
B_1 (r_h)=-1-\frac{21}{4} \bar{\eta}^2 \phi_0^2
+{\cal O}(\phi_0^4)\,,
\qquad 
B_2 (r_h)=1+\frac{3}{2} \bar{\eta}^2 \phi_0^2+{\cal O}(\phi_0^4)\,,
\label{Bap}
\ee
and hence $B_1(r_h)^2-B_2(r_h)=9\bar{\eta}^2 \phi_0^2+{\cal O}(\phi_0^4)$. 
Then, the angular Laplacian stability conditions (\ref{B12con}) are 
satisfied in this regime. 
For $\phi_0$ close to the order 1, the approximation (\ref{Bap}) 
loses its validity. 
In Sec.~\ref{paraspa1}, we will numerically find the parameter space in which all the linear stability conditions are satisfied.

\subsection{Parameter space consistent with the linear stability}
\label{paraspa1}

In Secs.~\ref{infexp} and \ref{hoexp}, we showed that there are model parameter spaces 
consistent with all the linear stability conditions both at large distances and the horizon. 
In order to confirm the stability in an intermediate regime between the horizon and spatial infinity, we numerically compute quantities associated with the angular Laplacian stability by using the background scalarized BH solutions. 
For given $\alpha$ and $\bar{\eta}(=\eta/r_h^2)$, we obtain the field value $\phi_0$ on the horizon leading to the boundary condition 
$\phi(\infty)=0$ at spatial infinity. Without loss of generality, 
we will focus on the case $\phi_0>0$.

In Fig.~\ref{EGBva}, we show an example for the radial 
dependence of $c_{\Omega+}^2$, $c_{\Omega-}^2$, and $c_{r2,{\rm even}}^2$ outside the horizon. 
For these values of $\alpha$ and $\bar{\eta}$, the squared angular propagation speeds computed from Eq.~(\ref{cosqeven}) are $c_{\Omega+}^2=1.30$ and $c_{\Omega-}^2=0.78$ on the horizon, which are in good agreement with their numerical values in Fig.~\ref{EGBva}. 
As $r$ increases from the horizon, both $c_{\Omega+}^2$ and $c_{\Omega-}^2$ continuously approach the asymptotic value 1. 
Since $c_{\Omega+}^2>0$ and $c_{\Omega-}^2>0$ throughout the horizon exterior, there are no angular Laplacian instabilities for even-parity perturbations. 
As we also observe in Fig.~\ref{EGBva}, $c_{r2,{\rm even}}^2$ 
is close to 1 outside the horizon. 
Numerically, we also confirmed that all the other linear stability conditions ${\cal G}>0$, ${\cal F}>0$, ${\cal H}>0$, and ${\cal K}>0$ are satisfied.

\begin{figure}[ht]
\begin{center}
\includegraphics[height=3.1in,width=3.0in]{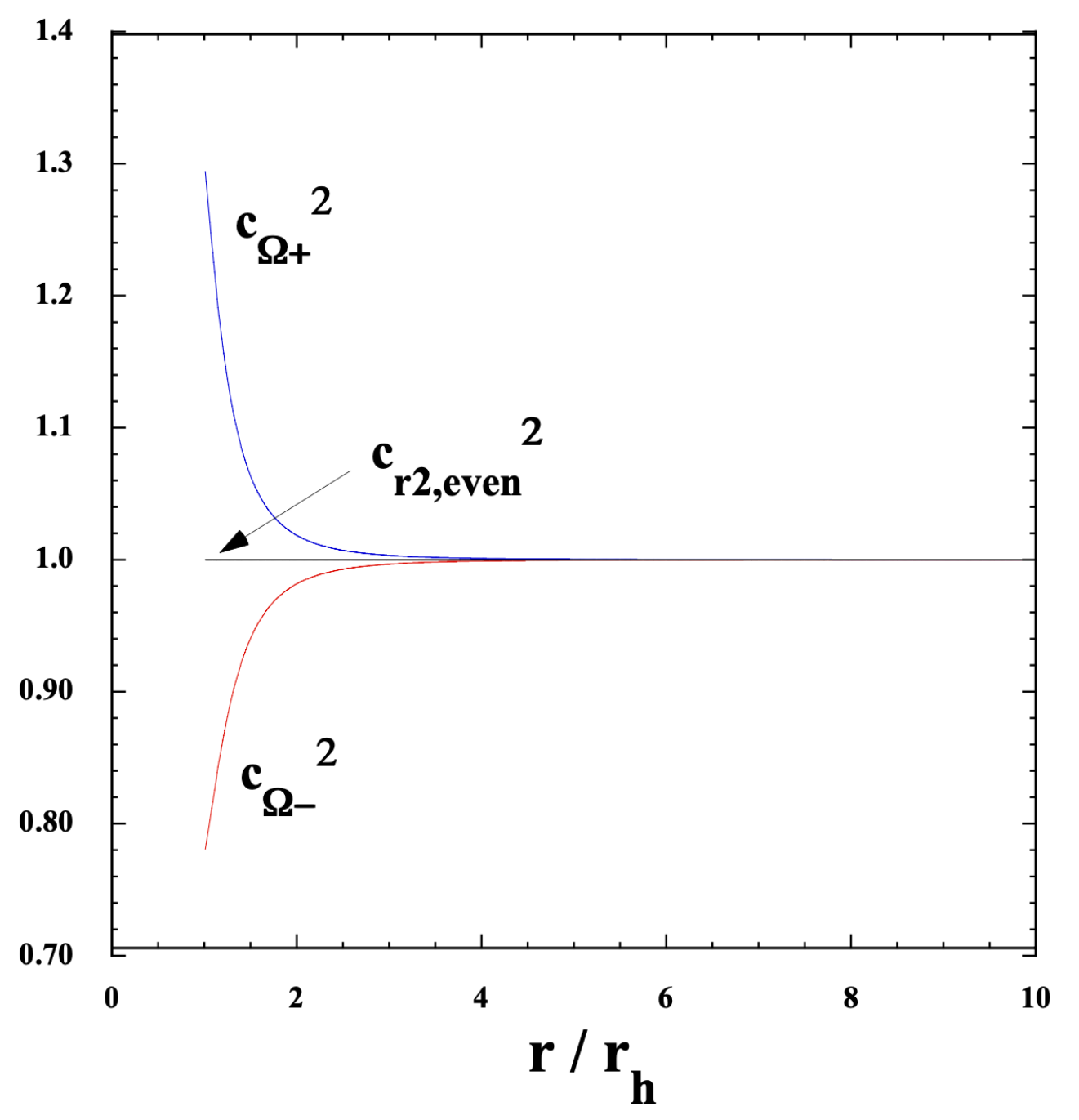}
\end{center}
\caption{We plot $c_{\Omega+}^2$, $c_{\Omega-}^2$, and $c_{r2,{\rm even}}^2$ versus $r/r_h$ for a scalarized BH with $\alpha=-10$ and $\bar{\eta}=1.0257$ in EsGB theories given by the coupling (\ref{model1}). 
In this case, the field value on the horizon realizing the asymptotic behavior $\phi(\infty)=0$ is $\phi_0=0.15$.
\label{EGBva} 
}
\end{figure}

\begin{figure}[ht]
\begin{center}
\includegraphics[height=2.5in,width=2.3in]{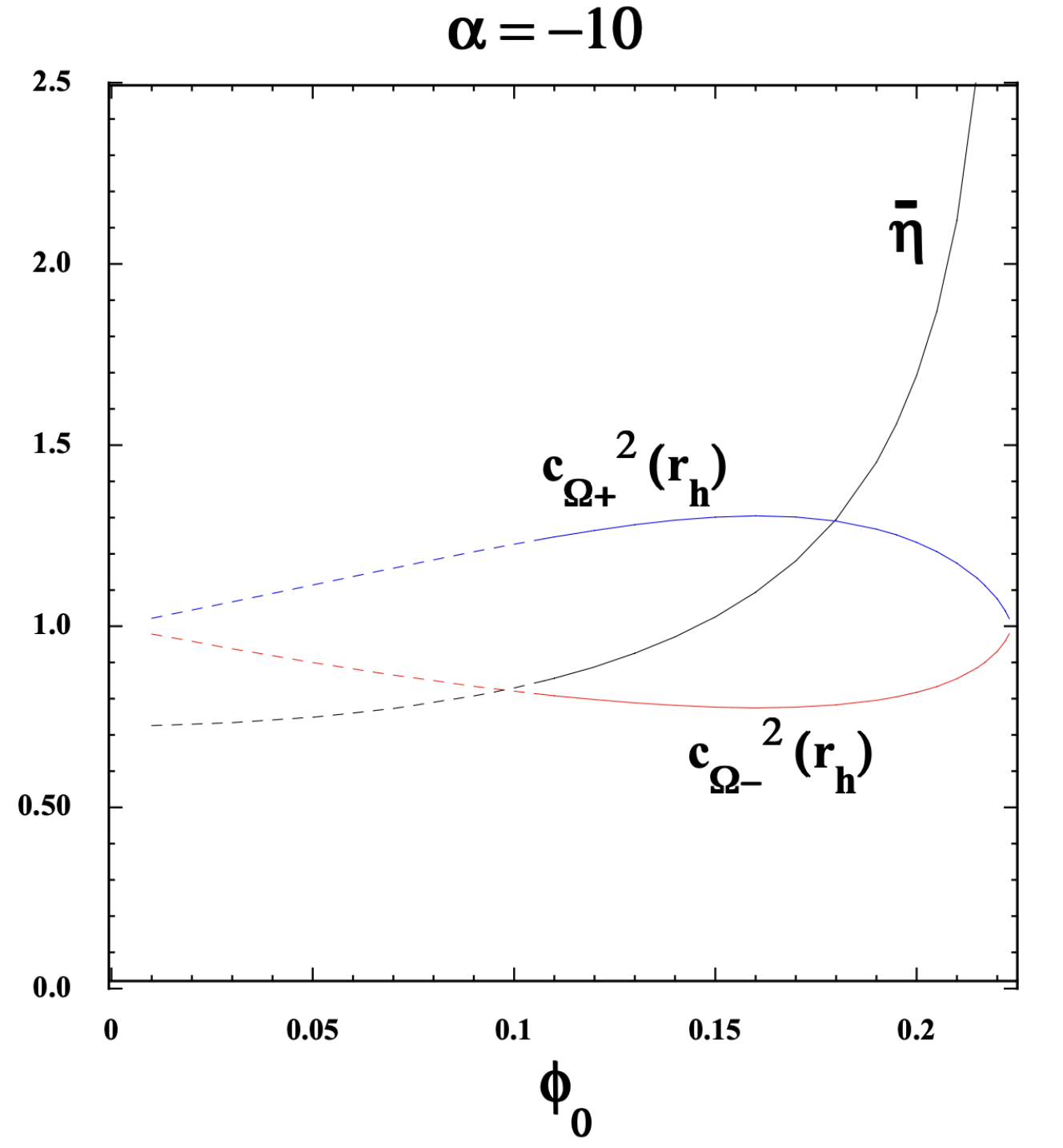}
\includegraphics[height=2.5in,width=2.3in]{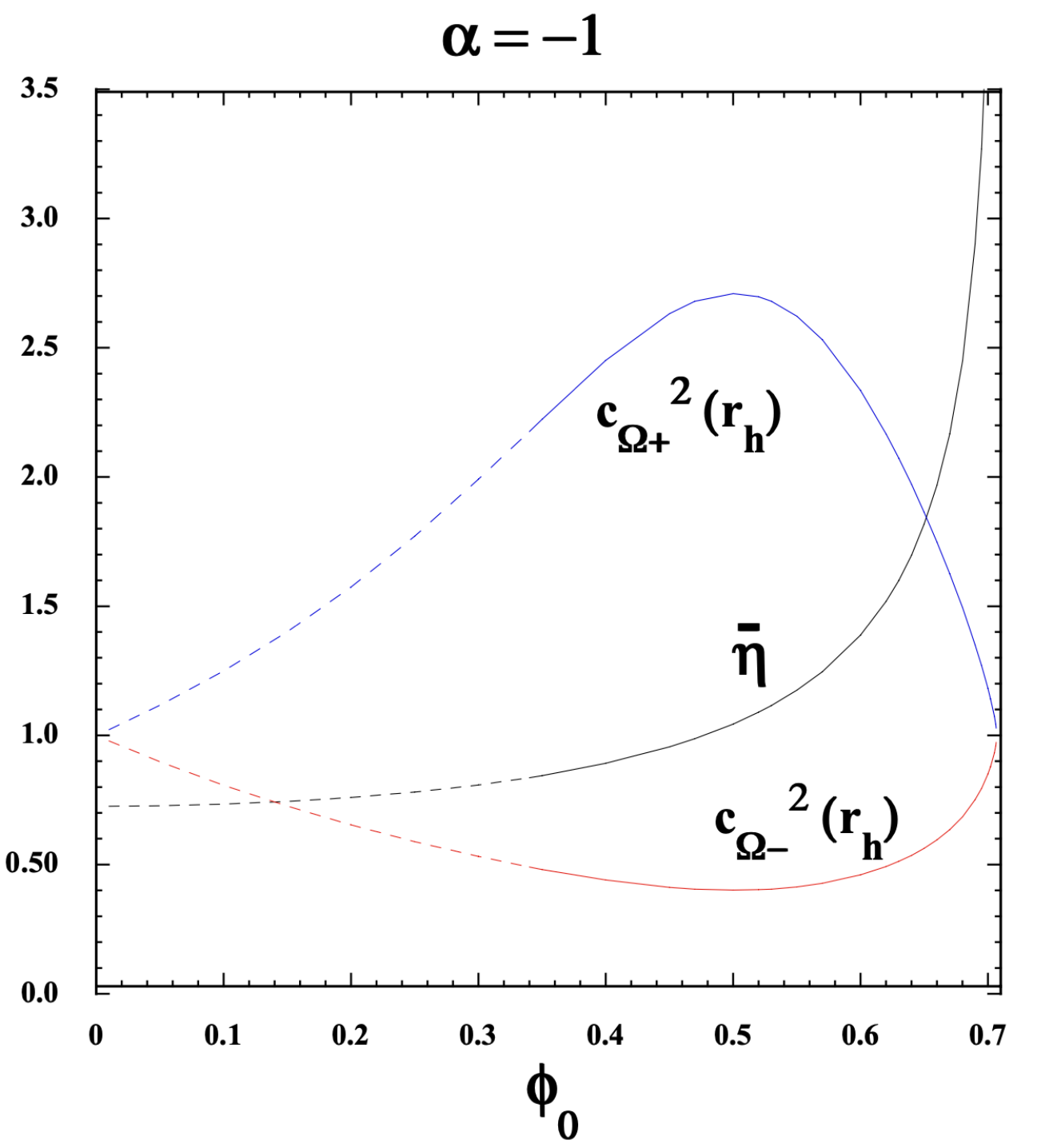}
\includegraphics[height=2.5in,width=2.3in]{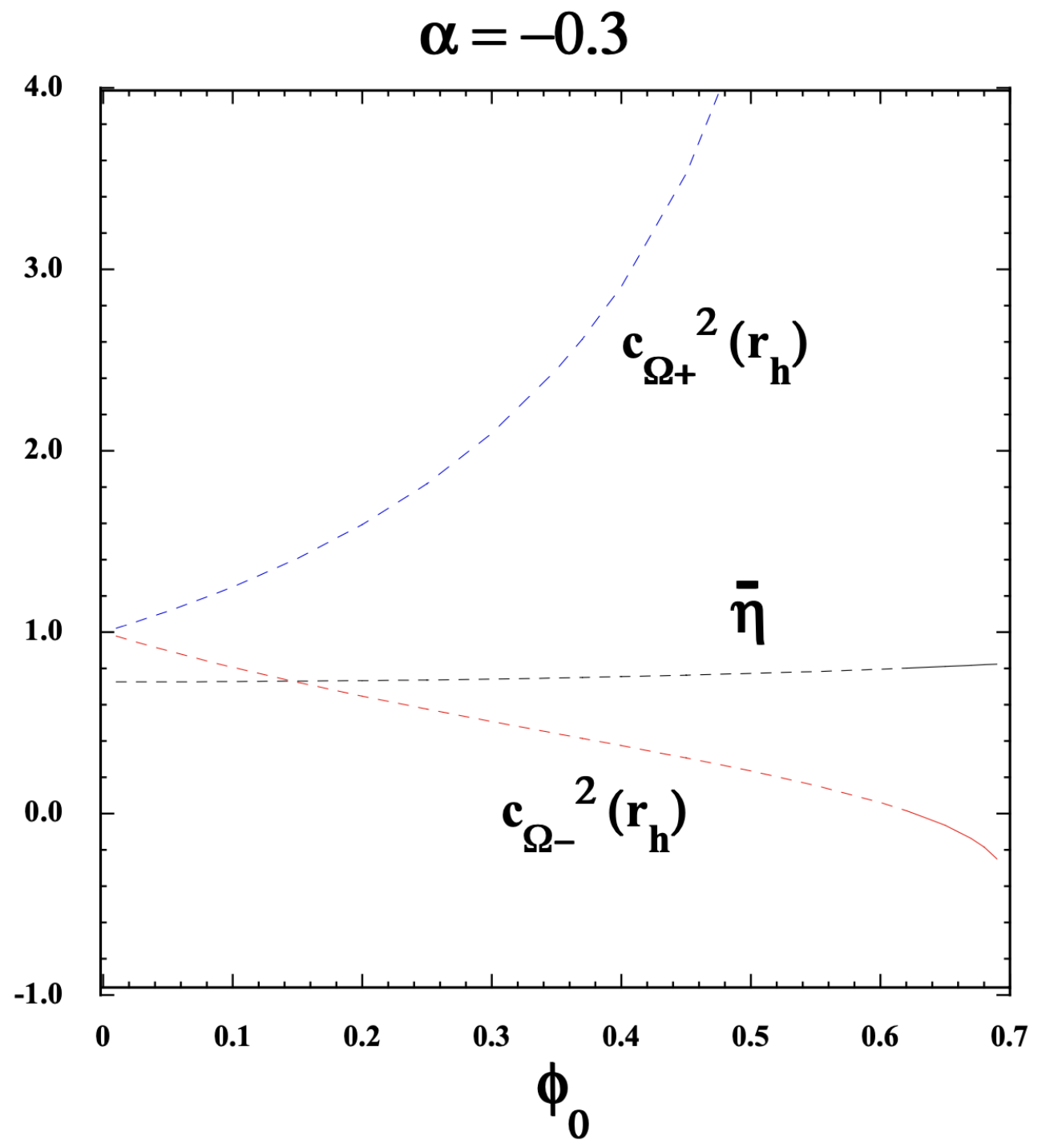}
\end{center}
\caption{The squared angular propagation speeds $c_{\Omega+}^2$ and $c_{\Omega-}^2$ on the horizon and $\bar{\eta}$ as functions of $\phi_0~(>0)$ in EsGB theories with the coupling (\ref{model1}). 
The left, middle, and right panels correspond to $\alpha=-10, -1, -0.3$, respectively. 
The dashed lines represent the regions excluded by the radial tachyonic instability, whereas the solid lines belong to the parameter range $-1/2<\alpha \phi_0^2<-0.1155$.\label{EGBsta} 
}
\end{figure}

We also carry out numerical simulations for the values of $\alpha$ and $\bar{\eta}$ different from those used in Fig.~\ref{EGBva}. 
The general property is that $c_{\Omega+}^2>1$ and $c_{\Omega-}^2<1$ 
outside the horizon. 
Then, for the angular Laplacian stability 
of even-parity perturbations, we only need to study  
whether $c_{\Omega-}^2$ is positive or not. 
Moreover, the deviation of $c_{\Omega-}^2$ from 1 is most significant 
at $r=r_h$ (see Fig.~\ref{EGBva}). 
Then, what determines the angular BH stability in the even-parity sector is the value $c_{\Omega-}^2$ on the horizon. 
In Fig.~\ref{EGBsta}, we plot $c_{\Omega-}^2(r_h)$ 
together with $c_{\Omega+}^2(r_h)$ and $\bar{\eta}$ as functions 
of $\phi_0$ by choosing three different values of $\alpha$.
For given $\alpha$ and $\phi_0$, the other parameter $\bar{\eta}$ is uniquely determined for the realization of scalarized BH solutions with the boundary condition $\phi (\infty)=0$. 
The dashed lines in Fig.~\ref{EGBsta}, which are 
outside the range (\ref{alphi}), are excluded by the radial tachyonic instability.

For $\alpha=-10$, $c_{\Omega-}^2 (r_h)$ has a minimum value 0.77 around $\phi_0=0.16$ and 
increases as a function of $\phi_0$ in the range $\phi_0>0.16$. 
For $\alpha=-1$, $c_{\Omega-}^2(r_h)$ exhibits a similar behavior, with the minimum value 0.40 around $\phi_0=0.50$.
The solid lines in Fig.~\ref{EGBsta} correspond to the range $-1/2<\alpha \phi_0^2<-0.1155$ (Eq.~\eqref{alphi}), in which the radial tachyonic instability is absent. 
As we increase $\phi_0$, the product $\alpha \phi_0^2$ approaches the lower limit $-1/2$. 
On using the expansion $\alpha \phi_0^2=-(1/2)(1-\epsilon)$, where $\epsilon$ is a small positive parameter close to 0, the squared angular propagation speeds can be estimated as
\be
c_{\Omega+}^2(r_h)=1+3\bar{\eta}\phi_0 \epsilon
+{\cal O}(\epsilon^2)\,,\qquad 
c_{\Omega-}^2(r_h)=1-3\bar{\eta}\phi_0 \epsilon
+{\cal O}(\epsilon^2)\,.
\ee
In the limit $\alpha \phi_0^2 \to -1/2$, both $c_{\Omega+}^2(r_h)$ and $c_{\Omega-}^2(r_h)$ approach 1. 
Indeed, this behavior can be seen in the left and middle panels 
of Fig.~\ref{EGBsta}. 
For $\alpha \lesssim -1$, we have $c_{\Omega-}^2(r_h)>0$ 
for the field value $\phi_0$ in the range $-1/2<\alpha \phi_0^2<-0.1155$. 
In such regions of the parameter space, we also confirmed that all the other linear stability conditions ${\cal G}>0$, ${\cal F}>0$, ${\cal H}>0$, ${\cal K}>0$, and $c_{r2,{\rm even}}^2>0$ are satisfied throughout the horizon exterior.

If $\alpha \gtrsim -1$, i.e., $|\alpha|$ less than the order 1, 
then we require larger values of $\phi_0$ to satisfy the condition $\alpha \phi_0^2<-0.1155$ as compared to the case $\alpha \lesssim -1$. 
As we observe in the right panel of Fig.~\ref{EGBsta}, which corresponds to $\alpha=-0.3$, the field value in the range $\phi_0<0.62$ is excluded by the radial tachyonic instability. For $\phi_0$ 
exceeding 0.62, $c_{\Omega-}^2(r_h)$ starts 
to enter the region $c_{\Omega-}^2(r_h)<0$.
When $\phi_0$ approaches the limit $\alpha \phi_0^2=-1/2$, we numerically find that it is difficult to realize scalarized hairy BH solutions with $\alpha \gtrsim -1$ due to the largeness of $\phi_0$ and the small variation 
of $\phi(r)$ around $r=r_h$. 
For $\alpha=-0.3$, $c_{\Omega-}^2(r_h)$ does not start to grow from negative values toward 1 with the increase of $\phi_0$. 
In the right panel of Fig.~\ref{EGBsta}, we see that there are almost no parameter spaces consistent with both angular Laplacian stability and radial tachyonic stability.

From the above discussion, so long as the coupling $\alpha$ and the horizon field value $\phi_0$ are in the ranges 
\be
\alpha \lesssim -1\,,\quad {\rm and} \quad
-\frac{1}{2}<\alpha \phi_0^2<-0.1155\,,
\ee
there are scalarized BH solutions compatible with all the linear stability conditions 
outside the horizon.

\section{BH linear stability in EsGBR theories}
\label{angsec2}

Finally, we study the angular Laplacian stability of scalarized BHs 
in EsGBR theories given by the couplings (\ref{model2}). 
We first consider the two asymptotic regimes far away from the horizon and 
around the horizon and then find the parameter space consistent with 
all the linear stability conditions.

\subsection{Region far away from the horizon}
\label{infexp2}

Imposing the boundary conditions $f \to 1$, $h \to 1$, and $\phi \to 0$ at spatial infinity, the background solutions to $f$, $h$, and $\phi$, 
which are expanded far away from the horizon ($r \gg r_h$), 
are given by 
\ba
f &=& 1-\frac{m}{r}+\frac{\beta q_s^2}{16 r^2}
-\frac{(3\beta-4)m q_s^2}{96 r^3}
-\frac{q_s^2 (3 \beta^3 q_s^2 -20 \beta^2 q_s^2 
+ 16\beta q_s^2 + 112 \beta m^2- 128 m^2)}
{3072 r^4}+{\cal O}(r^{-5})\,,\\
h &=& 1-\frac{m}{r}-\frac{(\beta-2) q_s^2}{8 r^2}
-\frac{(3\beta-4)m q_s^2}{32 r^3}
+\frac{q_s^2 (3\beta^3 q_s^2 -5 \beta^2 q_s^2
+4 \beta q_s^2 -56 \beta m^2+ 64 m^2)}
{768 r^4}+{\cal O}(r^{-5})\,,\\
\phi &=& \frac{q_s}{r} + \frac{m q_s}{2r^2} 
-\frac{q_s ( 3\beta^2 q_s^2 - 8\beta q_s^2 +16 q_s^2-128 m^2
)}{384r^3}-\frac{q_s m
 ( 9\beta^2 q_s^2 -36 \beta q_s^2 + 64q_s^2 - 192m^2
)}{768 r^4} 
+{\cal O}(r^{-5})\,,
\ea
where $m$ corresponds to twice the ADM mass, and 
$q_s$ is the scalar charge. 
Unlike the coupling $\alpha$ in EsGB theories, 
the nonminimal coupling constant $\beta$ appears at the 
order of $r^{-2}$ in $f$ and $h$ and of $r^{-3}$ in $\phi$. 
To obtain all the linear stability conditions correctly, 
we need to perform the expansions of $f$, $h$, and $\phi$ up to the order $r^{-9}$.

The quantities relevant to the stability of odd-parity 
perturbations are given by 
\be
{\cal G}=1-\frac{\beta q_s^2}{16 r^2}+{\cal O}(r^{-3})\,,\qquad
{\cal F}=1-\frac{\beta q_s^2}{16 r^2}+{\cal O}(r^{-3})\,,\qquad
{\cal H}=1-\frac{\beta q_s^2}{16 r^2}+{\cal O}(r^{-3})\,,
\ee
whose dominant terms are all positive. 
The leading-order contribution to ${\cal K}$ is the same as the first term in Eq.~(\ref{calK}), so that the no-ghost condition is satisfied at large distances. 
We also have $c_{r2,{\rm even}}^2=1+{\cal O}(r^{-8})$ and 
hence the radial Laplacian stability of even-parity perturbations 
is ensured. 

The quantities $B_1$ and $B_2$ have the following 
asymptotic behaviors 
\ba
& &
B_1 = -1+\frac{\eta q_s^2}{2r^4}
-\frac{\eta q_s^2[(15 \beta^2 - 16 \beta - 16) q_s^2 
+ 80 m^2]}{384 r^6}
-\frac{\eta q_s^2 m
[(15 \beta^2- 16 \beta - 32) q_s^2 
+ 128 m^2]}{384 r^7}
\nonumber \\
& &\qquad
+\frac{\eta q_s^2[( \beta \{320 +\beta [592 
+ 9 \beta (41\beta -108)] \} -512) q_s^4 
+ 96 \{ 176 + 5 \beta (2 - 5 \beta) \} q_s^2 m^2
- 51712 m^4-6144 \eta (10 q_s^2 + 99m^2
)]}{122880r^8} 
\nonumber \\
& &\qquad
+{\cal O}(r^{-9})\,,\label{B1b} \\
& &
B_2 = 1-\frac{\eta q_s^2}{r^4}
+\frac{\eta q_s^2 [(15 \beta^2-16 \beta-16)q_s^2
+80 m^2]}
{192 r^6}+\frac{\eta q_s^2 m
[(15 \beta^2 - 16 \beta - 32)q_s^2 + 128m^2
]}{192 r^7}
\nonumber \\
& &\qquad
-\frac{\eta q_s^2[( \beta \{320 +\beta [592 + 9 \beta (41\beta -108)] \} 
-512) q_s^4 + 96 \{ 176 + 5 \beta (2 - 5 \beta) \} q_s^2 m^2
- 51712 m^4
-6144 \eta (10 q_s^2 + 9m^2
)]}{61440 r^8} 
\nonumber \\
& &\qquad
+{\cal O}(r^{-9})\,.
\ea
This gives
\be
B_1^2-B_2=\frac{\eta^2 q_s^2 
(q_s^2+36 m^2
)}{4r^8}+{\cal O}(r^{-9})\,.
\label{B12b}
\ee
The leading-order contribution to $B_1^2-B_2$, which 
is the same as Eq.~(\ref{B12a}), is always positive. 
Since we have $B_1 \to -1$ and $B_2 \to 1$ as 
$r \to \infty$, the angular Laplacian stability 
conditions (\ref{B12con}) are consistently satisfied.

\subsection{Region in the vicinity of the horizon}
\label{hoexp2}

Around the horizon radius $r_h$, we expand $f$, $h$, $\phi$ 
in the same form as Eq.~(\ref{fhho}) and derive the coefficients 
$f_i$, $h_i$, $\phi_i$ by using the background equations of motion.  
While the coefficient $f_1$ is undetermined, 
$h_1$ and $\phi_1$ are given by 
\be
h_1=\frac{2}{r_h} \left[ 1+\sqrt{\frac{64 
+ [\beta (3 \beta-4 + 24\bar{\eta}) 
-384\bar{\eta}^2] \phi_0^2}
{64 + \beta (3 \beta -4 - 24\bar{\eta}) \phi_0^2}}
\right]^{-1}\,,\qquad 
\phi_1=\frac{12\bar{\eta} \phi_0 (\beta \phi_0^2-16)}
{64 + \beta (3 \beta -4 - 24\bar{\eta}) \phi_0^2}h_1\,.
\label{h1phi1}
\ee
To have a positive real value of $h_1$, 
we need the following condition 
\be
\left[ 64 + \beta (3 \beta -4 - 24\bar{\eta}) 
\phi_0^2 \right] \left\{ 64 
+ [\beta (3 \beta-4 + 24\bar{\eta}) 
-384\bar{\eta}^2] \phi_0^2 \right\}>0\,,
\ee
which holds for small $\phi_0$ close to 0.
Let us consider the regime in which the inequality
\be
64 + \beta (3 \beta -4 - 24\bar{\eta}) \phi_0^2>0
\ee
is satisfied. To realize a decreasing function of 
$|\phi(r)|$ around $r=r_h$ for $\bar{\eta}>0$, 
we require that 
\be
\beta \phi_0^2<16\,.
\ee
As $|\phi_0|$ approaches the upper limit $4/\sqrt{\beta}$, 
it tends to be difficult to obtain the scalarized BH solutions 
with the asymptotic behavior $\phi(\infty)=0$.

Let us consider the regime where $|\phi_0| \ll 1$, 
with $\beta$ and $\bar{\eta}$ at most of order 10. 
On using Eq.~(\ref{h1phi1}) together with the expansion 
around $\phi_0=0$, the quantities ${\cal G}$, 
${\cal F}$, and ${\cal H}$ at $r=r_h$ can 
be estimated as
\be
{\cal G}(r_h)=1+\left( \frac{3\bar{\eta}^2}{2} 
-\frac{\beta}{16} \right) \phi_0^2+{\cal O} (\phi_0^4)\,,
\qquad 
{\cal F}(r_h)=1+\left( \frac{3\bar{\eta}^2}{2} 
-\frac{\beta}{16} \right) \phi_0^2+{\cal O} (\phi_0^4)\,,
\qquad 
{\cal H}(r_h)=1-\frac{\beta}{16} \phi_0^2\,,
\ee
and hence the odd-parity linear stability is ensured for small $|\phi_0|$.
In the same regime, the leading-order contribution to ${\cal K}(r_h)$ is $9\bar{\eta}^2 \phi_0^2/4$, which is positive. 
The squared radial propagation speed $c_{r2,{\rm even}}^2$ is also equivalent to 1 on the horizon. 
Up to the order of $\phi_0^2$, the quantities $B_1(r_h)$ and $B_2(r_h)$ are of the same forms as those in Eq.~(\ref{Bap}). 
Hence, in the $\phi_0 \to 0$ limit, the angular Laplacian stability of even-parity perturbations is also ensured, with $c_{\Omega \pm}^2(r_h) \to 1$.

\subsection{Parameter space consistent with the linear stability}
\label{paraspa2}

Since the approximation of small $|\phi_0|$ used in Sec.~\ref{hoexp2} starts to lose its validity for $|\phi_0|$ exceeding the order 0.1, we need to compute the values of $B_1(r_h)$, $B_2(r_h)$, etc. without resorting to 
such an approximation. 
For given $\beta$ and $\bar{\eta}$, we search for field values $\phi_0$ that lead to the asymptotic behavior $\phi(\infty)=0$. 
We will consider the case $\phi_0>0$ 
without loss of generality.
We numerically compute the quantities 
${\cal G}$, ${\cal F}$, ${\cal H}$, ${\cal K}$, $c_{r2,{\rm even}}^2$, $c_{\Omega+}^2$, and $c_{\Omega-}^2$ from the horizon to 
a sufficiently large distance. 
As in EsGB theories given by the couplings (\ref{model1}), we find that the second squared angular propagation speed $c_{\Omega-}^2$ on the horizon determines the linear stability of scalarized BHs.
In other words, if the condition $c_{\Omega-}^2 (r_h)>0$ is satisfied, all the other linear stability conditions also hold outside the horizon. Hence we will focus on the behavior of $c_{\Omega-}^2 (r_h)$ in the following discussion.

\begin{figure}[ht]
\begin{center}
\includegraphics[height=2.5in,width=2.3in]{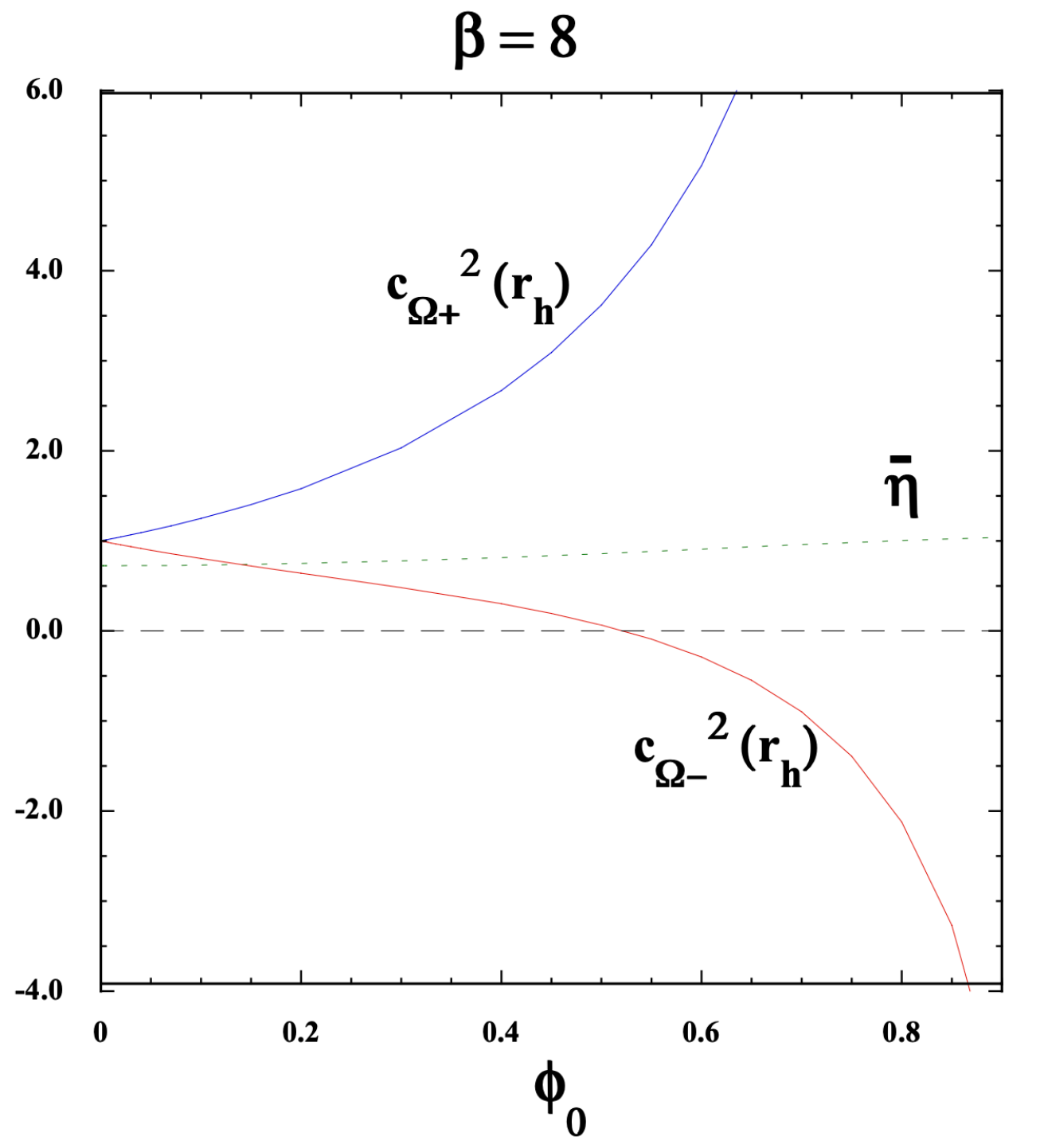}
\includegraphics[height=2.5in,width=2.3in]{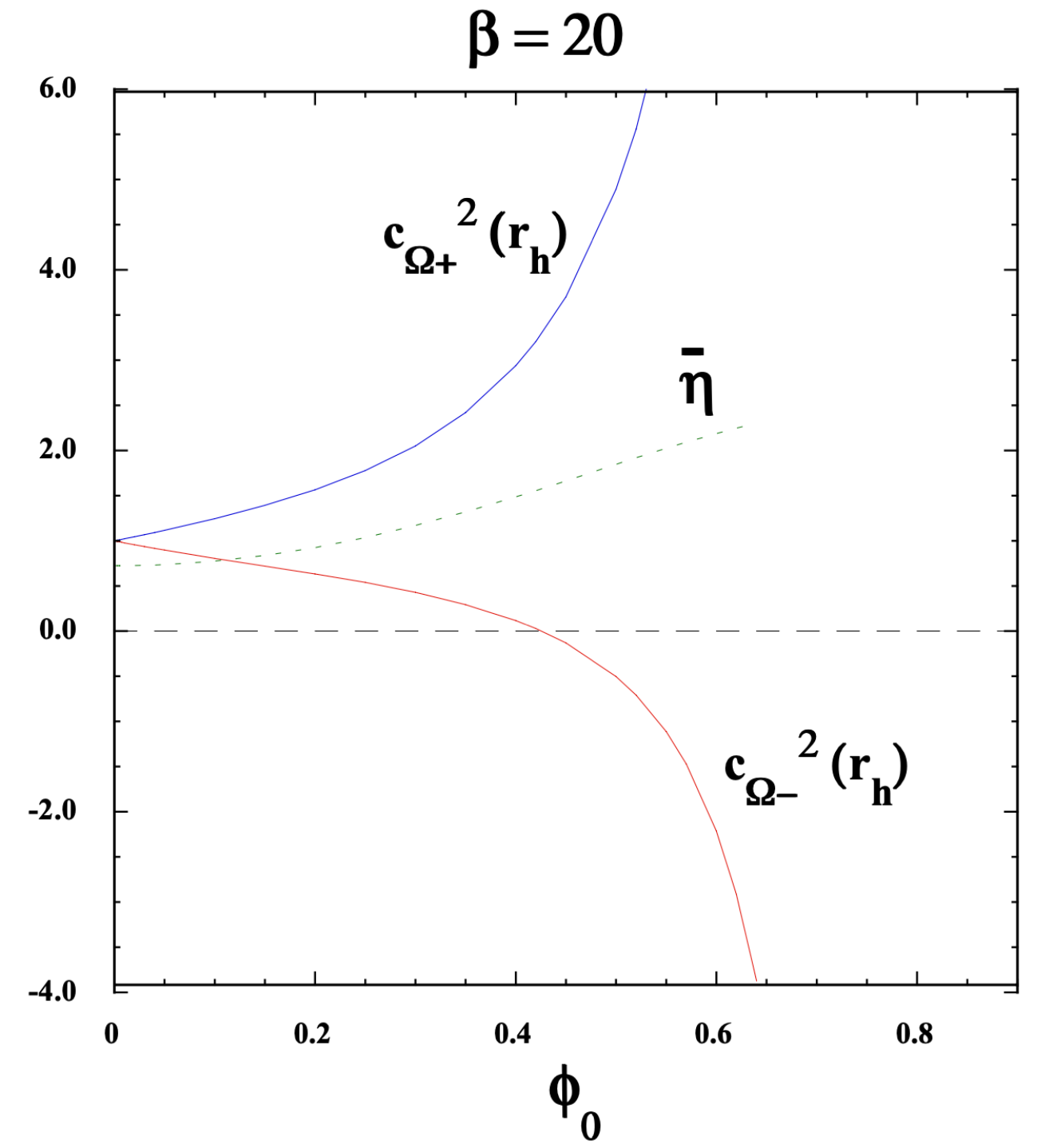}
\includegraphics[height=2.5in,width=2.3in]{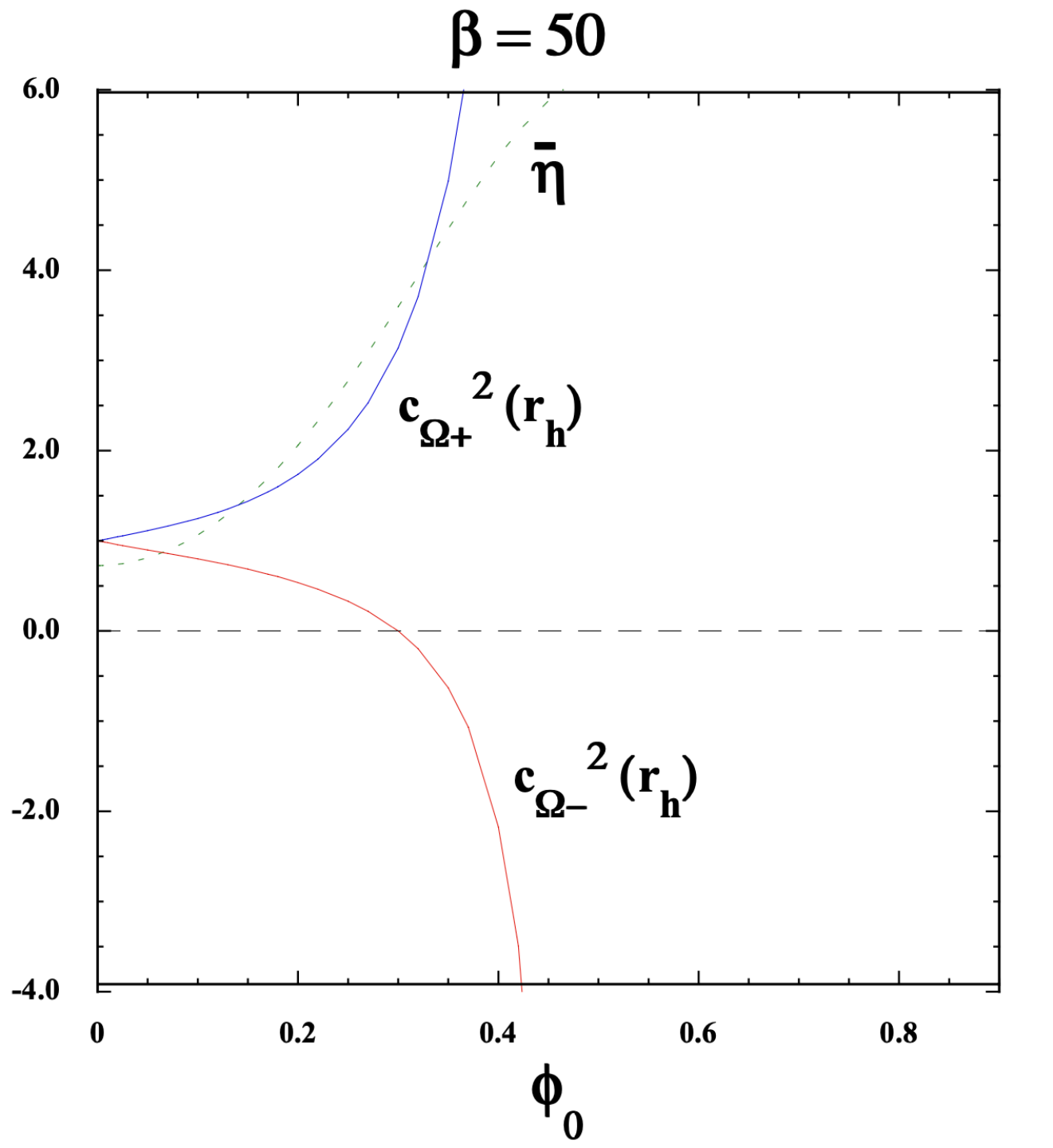}
\end{center}
\caption{The squared angular propagation speeds 
$c_{\Omega+}^2$ and $c_{\Omega-}^2$ at $r=r_h$ and 
$\bar{\eta}$ versus $\phi_0$ in EsGBR theories 
with the coupling (\ref{model2}). 
The left, middle, and right panels 
correspond to $\beta=8, 20, 50$, respectively. 
The field values $\phi_0$ that give $c_{\Omega-}^2(r_h)<0$ 
are excluded by the angular Laplacian instability of 
even-parity perturbations.
\label{EGBstaR} 
}
\end{figure}

In Fig.~\ref{EGBstaR}, we plot $c_{\Omega+}^2(r_h)$, $c_{\Omega-}^2(r_h)$, and $\bar{\eta}$ 
as functions of $\phi_0$ for three different values of $\beta$. 
While $c_{\Omega+}^2(r_h)$ is larger than 1 in all these cases, $c_{\Omega-}^2(r_h)$ decreases from the value close to 1 as $\phi_0$ increases. 
For $\beta=8$, $c_{\Omega-}^2(r_h)$ becomes negative in the region $\phi_0>0.52$. Even if $\phi_0$ increases toward the upper limit $\phi_0=4/\sqrt{\beta}$, neither $c_{\Omega+}^2(r_h)$ nor  $c_{\Omega-}^2(r_h)$ approaches 1. 
This property is different from that in EsGB theories given by the coupling (\ref{model1}). 
Instead, we find that $c_{\Omega-}^2(r_h)$ in EsGBR theories with the coupling (\ref{model2}) monotonically decreases with the increase of $\phi_0$.
Hence, for given values of $\beta$, there are upper bounds on $\phi_0$, below which the angular Laplacian stability condition $c_{\Omega-}^2(r_h)>0$ is satisfied. 

As we observe in Fig.~\ref{EGBstaR}, the maximum values of $\phi_0$ 
constrained from the condition $c_{\Omega-}^2(r_h)>0$ tend to be 
smaller for larger $\beta$. 
From the requirement $c_{\Omega-}^2(r_h)>0$, 
we obtain the following bounds 
\ba
& &
\phi_0<0.52\,,\quad {\rm for} \quad \beta=8\,,
\label{phi0con1}\\
& &
\phi_0<0.42\,,\quad {\rm for} \quad \beta=20\,,
\label{phi0con2}\\
& &
\phi_0<0.30\,,\quad {\rm for} \quad \beta=50\,,
\label{phi0con3}
\ea
respectively. As we discussed in Sec.~\ref{EsGBRbo}, the radial 
tachyonic instability of even-parity perturbations is absent 
for $\beta>5$ irrespective of the values of $\phi_0$. 
We also confirmed that, so long as the horizon field values are in the ranges (\ref{phi0con1})-(\ref{phi0con3}), 
all the other required conditions ${\cal G}>0$, ${\cal F}>0$, ${\cal H}>0$, ${\cal K}>0$, and $c_{r2,{\rm even}}^2>0$ hold outside the horizon. 
Thus, for $\beta>5$, there are viable parameter spaces of $\phi_0$ such as (\ref{phi0con1})-(\ref{phi0con3}) 
consistent with all the linear stability conditions. 

\section{Conclusions}
\label{consec}

In this paper, we studied the linear stability of scalarized BHs in scalar-tensor theories given by the action (\ref{action}). 
EsGB theories have the sGB coupling $\xi(\phi)R_{\rm GB}^2$ with $\kappa(\phi)=0$, whereas EsGBR theories possess the nonminimal 
coupling $\kappa (\phi)R$ besides the sGB coupling. 
Provided that $\xi(\phi)$ and $\kappa(\phi)$ are even power-law functions of $\phi$, there are in general a nonvanishing scalar-field branch ($\phi \neq 0$) and a GR branch ($\phi=0$). On the strong gravitational background like the vicinity of BHs, the GR branch can trigger a tachyonic instability 
toward the other scalarized branch 
under the condition $\xi_{,\phi \phi}>0$. 
A simple choice of the sGB coupling satisfying this inequality is $\xi(\phi)=\eta \phi^2/8$ with $\eta>0$. However, for the same coupling, the scalarized branch is subject to tachyonic instability against radial perturbations.
This problem can be circumvented by implementing higher-order terms in EsGB theories, e.g., $\xi(\phi)=(\eta/8) ( \phi^2+\alpha \phi^4 )$, or by taking into account the nonminimal coupling in EsGBR theories, e.g., $\xi(\phi)=\eta \phi^2/8$ and $\kappa(\phi)=-\beta \phi^2/16$. 

In Sec.~\ref{stabilitysec}, we revisited conditions for avoiding ghosts and Laplacian instabilities along the radial and angular directions on the background (\ref{spmetric}).
For the multipoles $l \geq 2$, there are three dynamical degrees of freedom: $\chi$ in the odd-parity sector, and 
$\psi$ and $\delta \phi$ in the even-parity sector.
For large radial and angular momentum modes, there are neither ghosts nor Laplacian instabilities under the seven conditions ${\cal G}>0$, ${\cal F}>0$, ${\cal H}>0$, ${\cal K}>0$, $c_{r2,{\rm even}}^2>0$, $c_{\Omega+}^2>0$, and $c_{\Omega-}^2>0$. In general, a scalarized BH solution has the largest 
scalar-field amplitude $|\phi_0|$ on the horizon. 
For the couplings $\xi(\phi)=(\eta/8) ( \phi^2+\alpha \phi^4)$ and $\kappa(\phi)=-\beta \phi^2/16$, which accommodate both the models in EsGB and EsGBR theories mentioned above, we showed that the five conditions ${\cal G}>0$, ${\cal F}>0$, ${\cal H}>0$, ${\cal K}>0$, and $c_{r2,{\rm even}}^2>0$ 
hold for small $\phi_0~(\ll 1)$ and the couplings $\bar{\eta}=\eta/r_h^2$ 
and $\beta$ in the ranges (\ref{phicon1}), (\ref{phicon2}), 
and (\ref{etacon}). 
The angular stability relevant to the squared propagation speeds $c_{\Omega \pm}^2$ in the even-parity sector requires more detailed analyses, which we performed in later Secs.~\ref{angsec1} and \ref{angsec2}.

In Sec.~\ref{radisec}, we addressed the radial tachyonic stability of scalarized BHs by considering the monopole perturbation $l=0$. 
In this case, the second-order action of odd-parity perturbations 
vanishes identically.
In the even-parity sector, we showed that the quadratic action reduces to that of a single propagating degree of freedom $\delta \phi$, see Eq.~(\ref{Lag3}) for the corresponding Lagrangian. We derived an effective potential $V_{\rm eff}(r)$ for $\delta \phi$ along the radial direction in the form (\ref{Veff}), which captures the backreaction of metric perturbations on the scalar-field perturbation. 
In EsGB theories, the leading-order contribution to $V_{\rm eff}(r)$ arises from the potential $U \simeq -\xi_{,\phi \phi}R_{{\rm GB}0}^2/h$, where the background GB invariant $R_{{\rm GB}0}^2$ does not vanish even in the vacuum.
In this case, we showed that the analysis of neglecting the backreaction of metric perturbations on $\delta \phi$ is a good approximation for estimating $V_{\rm eff}(r)$. 
For the sGB coupling (\ref{model1}), we confirmed that the full study of 
$V_{\rm eff}(r)$ without using such an approximation gives the condition 
$\alpha \phi_0^2<-0.1155$ \cite{Minamitsuji:2018xde} for the absence 
of radial tachyonic instability. 
In EsGBR theories, we found that it is important to implement the backreaction of metric perturbations on the effective potential of $\delta \phi$.
The radial stability is ensured for $\beta>5$ even for $|\phi_0|$ much smaller than 1, 
whose result is consist with the one obtained 
in Ref.~\cite{Antoniou:2022agj}.

In Sec.~\ref{angsec1}, we studied the linear stability of scalarized BHs in EsGB theories given by the coupling (\ref{model1}), by paying particular attention to the angular stability 
conditions of even-parity perturbations in the 
limit $l \gg 1$. 
On using the background solutions expanded at large distances, we first showed that 
there are no angular Laplacian instabilities associated with 
$c_{\Omega \pm}^2=-B_1 \pm \sqrt{B_1^2-B_2}$ far away from the horizon, see Eqs.~(\ref{B1a})-(\ref{B12a}). 
To realize a radially stable scalarized BH with a decreasing function of $|\phi(r)|$ around the horizon, we require that the product $\alpha \phi_0^2$ needs to be in the range (\ref{alphi}). 
We found that the tightest linear stability condition comes from the value $c_{\Omega-}^2$ 
on the horizon. 
For $\alpha \lesssim -1$, $c_{\Omega-}^2(r_h)$ can remain positive 
in the parameter range $-1/2<\alpha \phi_0^2<-0.1155$. 
On the other hand, for $\alpha \gtrsim -1$, the region with 
$c_{\Omega-}^2(r_h)<0$ starts to appear and the parameter space 
compatible with the radial stability 
tends to be smaller. 
Thus, so long as $\alpha \lesssim -1$ and $-1/2<\alpha \phi_0^2<-0.1155$, 
the scalarized BHs can be consistent with all the linear stability conditions. 

In Sec.~\ref{angsec2}, we addressed the linear stability issue for scalarized BHs in EsGBR theories given by the coupling (\ref{model2}). As we observe in Eqs.~(\ref{B1b})-(\ref{B12b}), the angular Laplacian instability far away from the horizon is absent in this case as well. The horizon field value should be in the range $\beta \phi_0^2<16$ to realize the scalarized BH with a decreasing function of $|\phi(r)|$ around $r=r_h$. 
As seen in Fig.~\ref{EGBstaR}, the second angular propagation speed squared on the horizon tends to enter the region $c_{\Omega-}^2(r_h)<0$ as $\phi_0$ increases toward the upper limit $4/\sqrt{\beta}$.
However, for each $\beta$ larger than the order 1, there are the regions of $\phi_0$ in which the condition $c_{\Omega-}^2(r_h)>0$ is satisfied, see Eqs.~(\ref{phi0con1})-(\ref{phi0con3}). 
Provided that $\beta>5$, these small values of $\phi_0$ are also consistent with the radial tachyonic stability as well as other linear stability conditions. 

We have thus shown that the scalarized BHs present in EsGB and EsGBR theories have the theoretically allowed parameter spaces in which all the linear stability conditions are satisfied throughout the horizon exterior. 
It will be of interest to explore how the observations of gravitational waveforms emitted from BH-BH binaries and of quasinormal modes during the ringdown phase put constraints on the model parameters further. 
One could also extend the present work to the case of spontaneously vectorized BHs \cite{Ramazanoglu:2017xbl,Annulli:2019fzq,Kase:2020yhw,Oliveira:2020dru,Garcia-Saenz:2021uyv,Silva:2021jya,Aoki:2022woy}. It will also be intriguing to see how the linear scalar-GB coupling (without breaking the shift symmetry) will imprint non-stealth modifications to the known stealth and approximately stealth BH solutions (see footnote~\ref{stealthsol} for a brief description of such solutions). 
These issues are left for future works.

\section*{Acknowledgements}

MM~was supported by the Portuguese national fund through the Funda\c{c}\~{a}o para a Ci\^encia e a Tecnologia in the scope of the framework of the Decree-Law 57/2016 of August 29, changed by Law 57/2017 of July 19, and the Centro de Astrof\'{\i}sica e Gravita\c c\~ao through the Project~No.~UIDB/00099/2020.
The work of SM was supported in part by the Grant-in-Aid for
Scientific Research Fund of the JSPS No.~24K07017 and the World
Premier International Research Center Initiative (WPI), MEXT, Japan.
ST was supported by the Grant-in-Aid for Scientific Research Fund of the JSPS No.~22K03642 and Waseda University Special Research Project No.~2023C-473. 

\appendix

\section{Coefficients in the second-order action 
of even-parity perturbations}
\label{AppA}

The coefficients that appear in the quantities 
$B_1$ and $B_2$ in Eqs.~(\ref{B1def}) and (\ref{B2def}) 
are given by 
\ba
\beta_1&=&\phi'^2a_4 e_4-2 \phi' c_4 a_4'
+ \left[  \left( {\frac {f'}{f}}+{\frac {h'}{h}}-\frac{2}{r} \right) a_4
+{\frac {\sqrt {fh}{\cal F}}{r}} \right] \phi' c_4+{\frac {f{\cal F} {\cal G}}{2r^2}}\,,
\notag\\
\beta_2&=& \left[ {\frac {\sqrt {fh}{\cal F}}{r^2} 
\left( 2 hr\phi'^2c_4+{\frac {rf' \phi'a_4}{f}}-\sqrt {fh}\phi'{\cal G} \right) }
-{\frac {2\sqrt {fh}\phi' a_4 {\cal G}}{r} 
\left( {\frac {{\cal G}'}{{\cal G}}}-{\frac {a_4'}{a_4}}+{\frac {f'}{f}}+\frac12 {\frac {h'}{h}}-\frac{1}{r} \right) } \right] a_1-{\frac {4{\cal F} {\cal G} fha_4}{r}}\,,\qquad\,\, \notag\\
\beta_3&=& \left( hc_4'-\frac{d_3}{2}+\frac12 h' c_4 \right) \phi' a_4
+ \left( {\frac {h'}{2h}}-\frac{1}{r}+{\frac {f'}{2f}}-{\frac {a_4'}{a_4}} \right)  
\left( {\frac {a_4 f'}{2f}}+2 h \phi'c_4+{\frac {\sqrt {fh}{\cal G}}{2r}} \right) a_4\notag\\
&&
+{\frac {\sqrt {fh}{\cal F}}{4r} \left( {\frac {f'}{f}}a_4+2 h \phi'c_4
+{\frac {3\sqrt {fh}{\cal G}}{r}} \right) }\,,
\ea
where 
\ba
\label{d3}
d_3 &=&
-{\frac {1}{r^2} \left( {\frac {2\phi''}{\phi'}}+{\frac {h'}{h}} \right) }a_1
+{\frac {2f}{ ( f' r-2 f ) \phi'} \left( 
{\frac {2\phi''}{h\phi' r}}
+ {\frac {{f'}^{2}}{f^2}}
- {\frac {f' h'}{fh}}
-{\frac {2f'}{fr}}
+{\frac {2h'}{hr}}
+ {\frac {h'}{h^2r}} \right) }a_4
\notag\\
&&
+{\frac {f' r-2 f}{fr}}\frac{\partial a_4}{\partial\phi}
+{\frac {\sqrt {f}}{\phi' \sqrt {h}r^2}}{\cal F}
-{\frac {{f}^{3/2}}{\sqrt {h} ( f' r-2 f ) \phi'} 
\left( {\frac {f'}{fr}}+{\frac {2\phi''}{\phi' r}}+{\frac {h'}{hr}}-\frac{2}{r^2} \right)}
{\cal G} \,,\notag\\
e_4&=&{\frac {1}{\phi'}}c_4'-\frac12 {\frac {f' }{f\phi'^2h}}a_4'
-\frac12 {\frac {\sqrt {f}}{\phi'^2\sqrt {h}r}}{\cal G}'
+{\frac {1}{h\phi' r^2} \left( {\frac {\phi''}{\phi'}}+\frac12 {\frac {h'}{h}} \right) }a_1
\notag\\
&&
+{\frac {1}{4h\phi'^2} \left[ {\frac { ( f' r-6 f ) f'}{f^2r}}
+\frac {h'  ( f' r+4 f ) }{hrf}
-{\frac {4f ( 2 \phi'' h+h' \phi' ) }{\phi' h^2r ( f' r-2 f ) }} \right] }a_4
+\frac12 {\frac {h'}{h\phi'}}c_4
-{\frac{f'r-2 f}{4r\phi'\sqrt{fh}}} {\cal H}_{,\phi}
\notag\\
&&+\frac12 {\frac {f' hr-f}{r^2\sqrt {f}\phi'^2{h}^{3/2}}}{\cal F}
+\frac12 {\frac {\sqrt {f}}{r\phi'^2{h}^{3/2}} 
\left[ {\frac {f ( 2 \phi'' h+h' \phi' ) }{h\phi'  ( f' r-2 f ) }}+\frac12 {\frac {2 f-f' hr}{fr}} \right] }{\cal G}
\,.
\ea
In the Lagrangian density (\ref{Lag}), the coefficients which are not presented in the main text are  
\ba
a_2 &=& 
\sqrt{fh} \left( \frac{a_1}{\sqrt{fh}} \right)' 
-\left( \frac{\phi''}{\phi'}-\frac{f'}{2f} \right)a_1
+\frac{r}{\phi'} \left( \frac{f'}{f}-\frac{h'}{h} 
\right)a_4\,,\qquad
c_1= -\frac{1}{fh}a_1\,,\notag\\
c_3&=& -\frac{1}{4\sqrt{fh}} \left[ 
2 ( rf' h+fh-f) \kappa_{,\phi}+h \phi' 
\{ r(rf'+4f) \kappa_{,\phi \phi}
-4f'(3h-1)\xi_{,\phi \phi} \} \right]
\,, \notag\\
c_6 &=& \frac{f'\phi'}{8f}a_1+\frac{f'r}{2f}a_4-\frac{1}{4}\phi' c_2
+\frac{1}{2}h \phi' rc_4+\frac{1}{4} \sqrt{fh} {\cal G}\,,\notag\\
e_1 &=& \frac{1}{\phi' fh} \left[ 
\left( \frac{f'}{f}+\frac{h'}{2h} \right) a_1
-2a_1'+a_2-2rh a_6 \right]\,, \qquad
e_2 =-\frac{1}{4}r^2 \sqrt{fh}\,,
\notag\\
e_3 &=& \frac{r^2 \sqrt{f}}{4 \sqrt{h}} 
\left( \xi_{,\phi \phi} R_{{\rm GB}0}^2
+\kappa_{,\phi \phi} R_0 \right)\,, 
\label{coeff1}
\ea
where 
\be
a_6=-\frac{\sqrt{f}}{\phi'\sqrt{h}} \left[ 
\frac{a_4'}{\sqrt{fh}}-\frac{a_4(hf'+fh')}{2(fh)^{3/2}}
+\frac{a_4}{r \sqrt{fh}}-\frac{{\cal F}}{2r} \right]\,.
\label{coeff2}
\ee

\bibliographystyle{mybibstyle}
\bibliography{bib}

\end{document}